\documentclass[12pt,preprint]{aastex}

\usepackage{graphicx}

\usepackage{enumerate}

\def\eg{{\em e.g.\ }}

\def\ie{{\em i.e.\ }}

\newcommand{\run}[1]{\texttt{HW-{#1}}}
\newcommand{\runt}{\texttt{TW}}
\newcommand{\indeg}[1]{{#1}^{\circ}}
\newcommand{\ds}{\displaystyle}
\newcommand{\myfrac}[2]{\frac{\ds {#1}}{\ds {#2}}}

\def\dtt{{\delta \theta_{T}}}
\def\phat{{\hat{\psi}}}

\newcommand{\od}[2]{\frac{d {#1}}{d {#2}}}
\newcommand{\meanphat}{\langle \hat\psi \rangle_\rho}
\newcommand{\meanpsi}{\langle\psi \rangle_\rho}



\begin{document}

\title{Relaxation of Warped Disks: the Case of Pure Hydrodynamics}

\author{Kareem A. Sorathia\altaffilmark{1}, Julian H. Krolik\altaffilmark{1}}

\and

\author{John F. Hawley\altaffilmark{2}}

\altaffiltext{1}{Department of Physics and Astronomy, Johns Hopkins University, Baltimore, MD 21218, USA} 

\altaffiltext{2}{Department of Astronomy, University of Virginia, Charlottesville VA 22904, USA}

\begin{abstract}
Orbiting disks may exhibit bends due to a misalignment between the angular momentum of the inner and outer regions of the disk.  We begin a systematic simulational inquiry into the physics of warped disks with the simplest case: the relaxation of an unforced warp under pure fluid dynamics, \ie with no internal stresses other than Reynolds stress.  We focus on the nonlinear regime in which the bend rate is large compared to the disk aspect ratio. When warps are nonlinear, strong radial pressure gradients drive transonic radial motions along the disk's top and bottom surfaces that efficiently mix angular momentum.  The resulting nonlinear decay rate of the warp increases with the warp rate and the warp width, but, at least in the parameter regime studied here, is independent of the sound speed.  The characteristic magnitude of the associated angular momentum fluxes likewise increases with both the local warp rate and the radial range over which the warp extends; it also increases with increasing sound speed, but more slowly than linearly.  The angular momentum fluxes respond to the warp rate after a delay that scales with the square-root of the time for sound waves to cross the radial extent of the warp.  These behaviors are at variance with a number of the assumptions commonly used in analytic models to describe linear warp dynamics.
\end{abstract}


\section{Introduction}

Astrophysical manifestations of warped accretion disks can occur at a myriad of scales in the Universe, requiring no more than a misalignment of angular momentum between the inner and outer regions of the disk.  The exact cause of this misalignment may vary.  One of the first analyses of misaligned fluid disk dynamics was in the context of a Kerr black hole whose spin is oblique to the rotation of the disk; the Lense-Thirring effect then forces disk precession \citep{bp75}.  Alternatively, a disk orbiting a proto-stellar binary at a tilt to the binary orbital plane may develop a warp due to the quadrupolar term in the gravitational potential \citep{papterq95,Lubow:2000}.  Even in the absence of an external torque, the orbital plane of the mass supply may change over time, \eg in the case of proto-stars, AGN, and the specific cases of SS433 and Her X-1.  Absent external effects entirely, disks may undergo self-induced warping due to their own radiation output \citep{pringle96}.  The presence of these warps is likely to have important consequences for the internal dynamics of the disk as well as observational significance.  

By definition, if a disk is warped, the orientation of orbital angular momentum in the disk changes with radius.  Thus, in the end, whether warps grow, decay, or mutate, their evolution is very largely dependent on how angular momentum in different directions is delivered to the disk (e.g., via the Lense-Thirring mechanism) and then moves through it (i.e., by the action of internal stresses).  Torques due to external causes can often be evaluated directly, but internal stresses are a much harder problem.  Although they are now known to be due to MHD turbulence stirred by the magneto-rotational instability \citep{bh98}, even in the case of flat disks it is difficult to quantify these stresses in detail without large-scale numerical simulations.  Little is known about the character of this sort of MHD turbulence in warped disks; it is quite possible that its nature changes in significant ways.  At the very least, in the fully three-dimensional context of warped disks, the number of interesting components in the stress tensor grows from the single one relevant to flat disks (the $r$-$\phi$ component) to all six independent components, particularly the other two off-diagonal ones.  The problem of describing the internal stresses is further complicated by the fact that disk bending induces radial motions because the bend misaligns the vertical pressure profiles of neighboring rings \citep{pp83}.  These contribute to the $r$-$z$ component of the Reynolds stress precisely because bending means that the disk midplane changes as a function of radius.  However, it is not easy to estimate the magnitude of this contribution to the stress because these radial motions can be limited by a variety of mechanisms, such as the work associated with fluid compression, changes in the pressure profiles over time, and the MHD stresses.

Because of these difficulties, the overwhelming majority of work in this field has instead described internal stresses in terms of the {\it ansatz} of \citet{ss73}.  Analyzing the radial structure of time-steady, unwarped, thin disks, they pointed out that the internal stress tensor has only one interesting element, the one responsible for conveying the component of angular momentum parallel to the disk's mean angular momentum in the radial direction (i.e., the $r$-$\phi$ component).  Arguing on the basis of dimensional analysis, they suggested that its magnitude should be $\alpha p$, where $p$ is the local pressure in the disk and $\alpha$ is a parameter thought to be of order or somewhat less than unity.  In subsequent treatments, this stress has often been modeled as being due to a phenomenological isotropic anomalous viscosity, so that it is proportional to the local fluid shear rate.  For example, \cite{pp83} developed a one-dimensional formalism for describing the evolution of linear disk warps in which they supposed that the internal stress {\it for all three off-diagonal components} was due to a phenomenological viscosity proportional to the local pressure and the appropriate fluid shear.  

It is convenient to divide this earlier work according to the severity of the warp.  \cite{np99} pointed out that the warp becomes nonlinear when the bending rate $d\theta/d \ln r > H/r$, where $\theta$ denotes the angle of the local orbital normal from a reference direction and $H$ is the disk scale height. Bends sharp enough to make $d\theta/d \ln r > H/r$ (as we will discuss later) induce transonic radial motions, so that the nonlinearity is in the sense that the perturbed velocities in the disk are comparable to or greater than the sound speed.  Because most accretion disks are expected to be quite thin, any substantial tilt between the disk's orientation at large radius and its orientation at smaller radius would be stretched over many decades of radius unless the warp is nonlinear. For this reason, nonlinear warps are an important topic.

Nonetheless, equations are always much more tractable in the linear regime, so a large part of the analytic effort devoted to warps in disks has been restricted to the limit of gentler warps.  This linear regime is generally further subdivided into two limiting cases: the diffusive limit, in which $\alpha > H/r$; and its opposite, the bending wave limit.  In both cases, the warp still drives radial motions, but they are subsonic.  They are, however, generally treated differently.  In the diffusive case, the associated Reynolds stress is seen as driving warp relaxation, but the isotropic anomalous viscosity limits its amplitude.  The end-result is that warp relaxation is modeled by a diffusion equation whose diffusion coefficient $\alpha_2 \propto \alpha^{-1}$ \citep{p92}.  In the long-wavelength bending wave limit, rather than causing warp relaxation, the radial motions are thought to create a circulatory motion in the poloidal plane which drives a propagating wave with speed $\simeq c_{s0}/2$, where $c_{s0}$ is the isothermal sound speed.  The role of isotropic anomalous viscosity is then restricted to a slow damping of the bending wave \citep{pl95,demianski97,ivanov:97,Lubow:2000,Lubow:2002}.  In fact, it is this estimate of the bending wave damping rate that underlies the distinction between these two regimes according to the ratio $\alpha/(H/r)$ \citep{pl95}.

By separating the dynamics in terms of velocity scale (orbital vs. transonic radial motions vs. net inflow), performing an asymptotic expansion in $H/r$, and then integrating over spherical shells, \cite{o99} was able to transform a nonlinear description of the fluid motions into a 1-d equation for the radial evolution of the disk's local orientation.  In the course of this transformation, he derived expressions for the $\alpha_2$ parameter.  However, extending this formalism into the bending wave regime has proved difficult.  One expression of this difficulty is the fact that no solution for $\alpha_2$ can be found when $\alpha = 0$, the epicyclic frequency $\kappa$ is less than the orbital frequency $\Omega$, and the bending rate $d\theta/\ln r \gtrsim 0.2$ \citep{o99}.  Unfortunately, this range of parameters is, in fact, generic for nonlinear bending waves in the absence of internal stresses because hydrostatic equilibrium with non-zero pressure {\it always} makes $\kappa < \Omega$.

Given the obstacles to analytic work in nonlinear fluid dynamics, many have also turned to numerical simulation.  The overwhelming majority of simulations performed thus far has used the smoothed particle hydrodynamics (SPH) method including an isotropic anomalous viscosity.  These simulations have explored the dynamics of warp evolution for warps produced by a binary potential \citep{larwood:96}, warps produced through a Lense-Thirring torque \citep{np00}, and the relaxation of an unforced warp \citep{np99,lp07,lp10}.  Their results are generally consistent with the linear theory (likewise including an isotropic anomalous viscosity) in both the diffusive and bending wave regimes.  Some nonlinear effects have also been explored.

However, there are still a number of ways in which our understanding of this phenomenon remains unsatisfactory.  In particular, nearly all of the extant results depend on a model for the disk's internal stresses that does not correspond directly to the actual physical mechanism.  Unlike the $\alpha$-model of stress, MHD stresses are not necessarily linked to the local pressure.  Unlike the stress due to an isotropic shear viscosity, it is the time-derivative of the Maxwell stress that is related to the shear, not the stress itself.  Moreover, the nature of the relationship of stress to shear is different from that of shear viscosity, is highly anisotropic, incorporates proportionality factors quadratic in the magnetic field itself, and depends on gradients of the magnetic field in addition to gradients of the fluid velocity.  Thus, it is unclear to what degree either an isotropic phenomenological viscosity or a diffusivity of the $\alpha_2$ variety mimics the effects of these turbulent Maxwell stresses.

In order to answer these questions, we are embarking on a program in which we will approach the problem of internal stresses in warped disks from a different point of view.  We will consider only effects derivable directly from hydrodynamics and magnetohydrodynamics.  To be able to identify which mechanisms are responsible for which effects, we will introduce them one by one.  In this paper, we wish to study the effects of the fluid motions induced by the warp in isolation---without MHD and also without any phenomenological viscosity to artificially limit them.  Thus, this work might be thought of as studying the nonlinear bending wave regime because $\alpha \equiv 0$; put another way, numerical techniques permit study of exactly the interesting parameter regime in which the extant nonlinear analytic theory fails.  With the data from numerical hydrodynamics simulations, we can measure the Reynolds stresses driven by warps and explore their dependence on initial warp strength and physical parameters of the disk such as sound speed.  We will introduce MHD effects in a later paper now in preparation.

The plan of the paper is as follows: Section 2 describes our numerical method and the properties of the warped disks we simulate; Section 3 presents an overview of how disks with nonlinear warps relax when only hydrodynamic mechanisms are active; Section 4 analyzes the internal stresses in greater detail; and Section 5 summarizes our results and places them in the context of previous work.  In Appendix A we describe how we tested our methods for numerical dissipation.
%

\section{Warped Disk Model}
\label{sec:warpmod}

\subsection{Equations, initial and boundary conditions, and grid definition}

We study the hydrodynamical mechanisms involved in warp relaxation through simulations conducted using the {\it Zeus} algorithm \citep{zeusold}.  We use an implementation of {\it Zeus} that solves the three-dimensional equations of hydrodynamics in coordinates that have a diagonal three-metric, $g_{ii}$.  In this formulation, the equation of mass conservation is
\begin{equation}\label{masscon}
{\partial\rho \over \partial t} + {1\over\sqrt\gamma}~\partial_i
\left({\sqrt\gamma}\rho u^i \right) = 0
\end{equation}
where $\rho$ is the density, $\gamma$ is the determinant of the three-metric, and $u^i$ is the contravariant velocity component.  The $j$ component of momentum density is defined in terms of the covariant momentum density, $\rho w_j$, so that the evolution equation for the momentum density is written as
\begin{equation}\label{euler}
{\partial{\rho w_j}\over \partial t} + {1\over\sqrt\gamma}~\partial_i
\left({\sqrt\gamma} \rho w_j u^i\right) + \partial_j\left(P+Q_{jj}\right)
- {\rho u^k u^k \over 2}\partial_j g_{kk}  - \rho \partial_j \Phi = 0
\end{equation}
where $P$ is the pressure, and $\Phi$ is the gravitational potential.  The derivative of the metric components accounts for the inertial forces associated with a choice of coordinate system (e.g., centrifugal and Coriolis forces).   The symbol $Q$ denotes the stress tensor associated with the artificial bulk viscosity required for proper treatment of shocks; it is diagonal, and $Q_{jj}$ is the element associated with direction $j$.  We use an ideal gas equation of state, $P = \left(\Gamma-1\right)e$, where $e =\rho\epsilon$ is the internal energy and  $\epsilon$ is the specific internal energy. The internal energy equation is
\begin{equation}\label{intenergy}
{\partial e \over \partial_t} =
{1\over\sqrt\gamma}\partial_i\left({\sqrt\gamma}e
u^i\right) + {\left(P +Q_{ii}\right)\over\sqrt\gamma}\partial_i
\left({\sqrt\gamma} u^i \right) = 0
\end{equation}
The magnitude of the artificial viscous stress $Q_{ii}$ is based on the convergence of the physical velocity components, $v^{i}$ as in \cite{zeusold}, 
\begin{equation}\label{Qdef}
Q_i=\left(n_s \Delta x^i\right)^2\rho \left(\partial_i v^{i}\right)^2, 
\end{equation} 
when $\partial_i v^{i} < 0$ (no sum over $i$); $Q=0$ elsewhere.  The constant $n_s$ determines the number of zones over which the artificial viscosity spreads out a shock.  In {\it Zeus} applications this is typically $n_s = 2$, which we use here.  This numerical viscosity thermalizes kinetic energy lost in the shocks by increasing the entropy of the gas; by following entropy generation we can thus determine when shocks are an important dynamical element of an evolution.  The artificial viscosity also reduces zone-to-zone oscillations that would be excited by such shocks and improves code stability.  By design, the artificial viscosity has negligible (or zero) magnitude except in strongly compressive regions, nor does it create any shear viscosity.

Because we are studying warped disks, there is no way to construct a single system of coordinates in which the orbital velocity consistently points parallel to a grid coordinate.  Consequently, greater numerical diffusion can be expected than in simulations of flat disks.  On the basis of the tests we discuss at greater length in Appendix~\ref{sec:num}, we decided that spherical coordinates were the best choice for conducting our simulations.  In spherical geometry, even when disks are tilted by as much as $\indeg{30}$ relative to the coordinates, numerical artifacts are held to a very low level when the gridscale gives a resolution of at least $\sim 8$ Zones Per vertical scale Height (ZPH).  For all our simulations, the computational domain is $(r,\theta,\phi) \in [1,19] \times [0.05,0.95]\pi \times [0,2\pi]$.  The unit of length is arbitrary because Newtonian gravity has no characteristic scale.  The size of the radial cells is logarithmically graded, with $\Delta r$ increasing outwards, and the azimuthal variable, $\phi$, is uniformly spaced.  Both $\Delta r/r$ and $\Delta \phi$ are set equal to $\Delta \theta$ so that the spatial resolution is isotropic. For the HW series of simulations (defined below), the polar angle, $\theta$, is also uniformly spaced; however the increased computational cost would make this uniform spacing infeasible for the thinner TW simulation.  Instead, the polar angle is spaced using a polynomial spacing (Equation 6 of \citet{nkh10}, with $\xi = 0.7$ and $n = 11$) to ensure adequate resolution near the midplane while accepting lower resolutions at higher altitudes.  For this simulation, $\Delta \phi = 2 \Delta \theta (\theta=\pi/2) = 2\Delta r/r$.  The resolutions used for the simulations presented here have $(N_{r},N_{\theta},N_{\phi}) = (128,128,192)$ and $(N_{r},N_{\theta},N_{\phi}) = (256,128,240)$ grid cells for the HW and TW runs, respectively.  These simulations are run using outflow boundary conditions in the radial and polar coordinates and a periodic boundary condition in the azimuthal direction.  The lack of a magnetic field in these simulations implies that there will be no consistent stress to drive inward flow, and thus there is little interaction between the disk and the boundary.    

To avoid both dynamical transients associated with the disk boundaries and numerical effects associated with the boundary conditions of the simulation domain, we use as our initial condition a state derived from an exactly hydrostatic torus \citep{2000ApJ...528..462H}.  Warps, of course, introduce non-hydrostatic features, so our simulations do not begin from an equilibrium.  Our initial conditions are constructed in two steps: first we lay down a density, pressure, and velocity distribution consistent with a flat hydrostatic torus; then we apply a systematic warp.  The result is that the angle $\theta_T$ between the local orbital normal (the direction of the total angular momentum at a particular spherical radius $r$) and the $z$-axis is a specified function of $r$.  We define the $x$-axis so that the tilt is in the $x$-$z$ plane.

A hydrostatic torus is fully determined by five parameters: the shear parameter $q$ determined by the non-Keplerian orbital velocity profile ($\Omega \propto R^{-q}$); the adiabatic index $\Gamma$; the radius of the inner edge, $R_{in}$; the radius of maximum pressure, $R_{M}$; and the maximum density, $\rho(R=R_{M},z=0) = \rho_{M}$.  The units of density are arbitrary because the gas is not self-gravitating.  Because of its axisymmetry, the properties of such a torus are functions of only $R$ and $z$ (cylindrical radius and height from the midplane), \ie $\rho(R,z)$, $e(R,z)$, and $v_{\phi} (R) = R\Omega \hat{\phi}$.  These dependences can easily be translated to the spherical coordinates of the simulation by the usual relations.  Its midplane is, of course, located at $z=0$.

In the work presented here we use two different initial tori.  For both, $\Gamma = 5/3$, $R_{in} = 2$, $R_{M} = 4$, and $\rho_{M} = 100$.  However, they differ in their values of $q$.  For our fiducial model, designated (H)ydrodynamic (W)arped disk (HW), $q=1.6$; in our alternate model, called (T)hin (W)arped disk (TW), $q=1.52$. As a result, they differ in their typical sound speeds and aspect ratios.  In the HW torus, $H/r \approx 1/4$ throughout the body of the disk, while the TW torus is more nearly Keplerian and therefore also has less pressure support, so that $H/r \approx 1/10$.

To warp the torus, we first choose a function $\theta_T(r)$.  In order to isolate the warp in the middle of the disk and ensure continuous variation, we use functions of the form
\begin{equation}
  \label{eqn:thetaprof}
   \theta_{T} = \left\{
    \begin{array}{ll}
             0 & : r \leq r_{c}-L_{W} ,\\
             A \ln(r) + B & : |r-r_{c}| \leq 2 L_{W} \\
      	    \theta_{F} & : r \geq r_{c}+L_{W}
    \end{array}
  \right.
\end{equation} 
where $\theta_{F}$ is the angle defining the orbital inclination of the outer disk, and the coefficients $A$ and $B$ are chosen so that $\theta_{T}$(r) is continuous. The midpoint of the bend $r_{c}$ is set to coincide with the radius of the spherical shell bounding half the total angular momentum of the system.   The central radius $r_{c} = 9$ for the HW series and $r_{c} = 8.3$ for the TW simulation. 

With this function specified, we define a warped cylindrical coordinate system ($R^\prime,\phi^\prime ,z^\prime$) related to the original cylindrical coordinate system through rotating a Cartesian coordinate scheme by $\theta_{T}(r)$ about the $y$-axis and then using the usual definition (i.e., $R^\prime \equiv ({x^\prime}^2 + {y^\prime}^2)^{1/2}$, $\phi^\prime \equiv \tan^{-1}(y^\prime/x^\prime)$).  The data defining the original hydrostatic torus are then remapped onto this new coordinate system by identifying points: $\rho(R^\prime,z^\prime) = \rho(R,z)$, etc.  The magnitude of the orbital velocity is preserved in the midplane, but its direction is, of course, made parallel to $\hat \phi^\prime$.  Off the midplane, the warped disk is constrained to orbit on cylinders.  This warped coordinate system is used only for construction of the initial condition; the simulation itself is conducted in spherical coordinates.

\subsection{Quantifying the warp: $\hat\psi$}\label{sec:psihat}

To complete the definition of a warped disk, we must choose the parameters $A$, $B$, and $\theta_F$.  We do so by characterizing the warp in terms of the dimensionless quantity
\begin{equation}
\psi(r) \equiv \od{\theta_{T}(r)}{\ln r} .
\label{eqn:psi}
\end{equation}
The functional form we have chosen results in $\psi(r)$ being piecewise constant.  Note, however, that our use of $\psi$ in this context differs slightly from the standard definition for $\psi$ ($\equiv r | \partial \vec{\ell}/\partial r|$, with $\vec{\ell}$ a unit vector pointing in the direction of the local angular momentum), in that Equation~\ref{eqn:psi} is a signed quantity and measures only rotation about the $y-$axis.  In the case of negligible $\ell_{y}$, our quantity agrees in magnitude with the standard definition.

As first remarked by \citet{np99}, the linear regime in a disk with $H/r \ll 1$ is defined by $\psi \ll H/r$, suggesting that $\hat\psi \equiv \psi/(H/r)$ is also an important measure of disk warp.
To see why this is so, we will use analytic estimates here and reinforce these points quantitatively when we describe our numerical results.

The fundamental reason why $\hat\psi > 1$ marks the onset of nonlinearity is that this is the criterion for the vertical displacement across a radial separation $\sim r$ to be larger than a scale height $H$.   When this is so, the warp creates a radial pressure contrast at fixed height from the midplane that is order unity.  As a result, fluid forces in the radial direction are no longer small perturbations to the flow.

\begin{figure}
\begin{center}
\includegraphics[width=\textwidth]{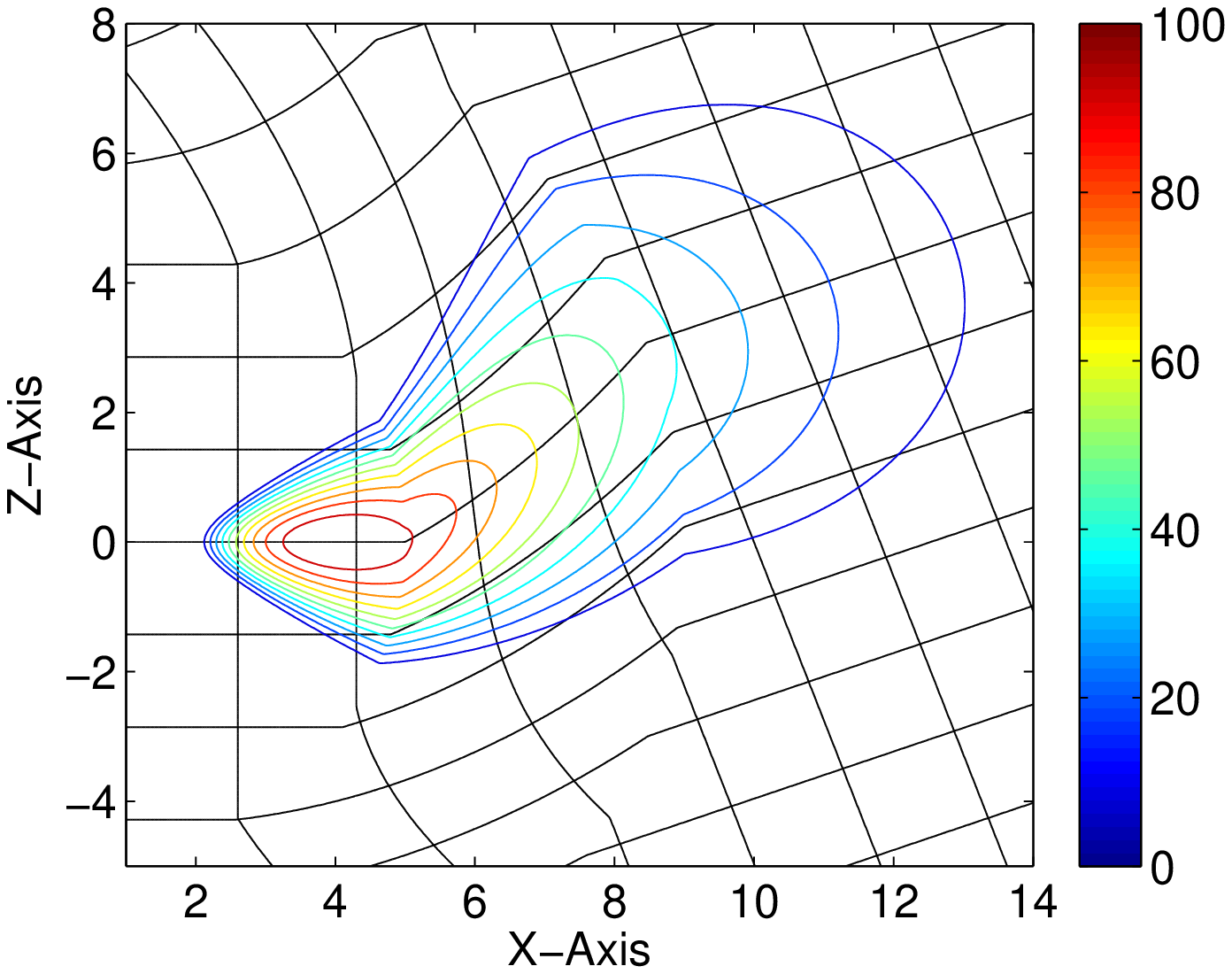}
\caption{Density contours of a slice in the $x-z$ plane through a warped disk (color curves) shown relative to two coordinate systems.  The axes are marked in terms of Cartesian coordinates $(x,z)$, while the black lines show the grid of the warped cylindrical coordinate system $(R',z')$ in this plane.  Note that the $(R',z')$ coordinates are not orthogonal in the warped region of the disk due to the non-zero derivative of $\theta_{T}$, however a local orthonormal basis can be defined.  The lack of orthogonality does not concern us here as these coordinates are merely used to define our initial conditions.}
\label{fig:warpcoord}
\end{center}
\end{figure}

In addition, when $\hat\psi > 1$ across a radial extent $\sim r$ or more, the elongation of the pressure contours in the warped region becomes significantly offset from the radial direction (see Fig.~\ref{fig:warpcoord}).  That is, the plane of the pressure distribution does not coincide with the local orbital plane.  In a thin disk, the local pressure gradient lies very close to the normal to the equatorial plane, but the pressure gradient due to a warp tilts the pressure gradient oblique to the orbital normal.  When $\hat\psi > 1$, this obliquity becomes sizable.   Consequently, the matter cannot remain hydrostatic even in the rotating frame.  Moreover, because the pressure gradient (whose natural length scale is $H$) is not balanced by gravity, the fluid velocity it induces can be large.  Integrating the acceleration due to the pressure gradient over an orbital period leads to transonic radial velocities:
\begin{equation}\label{eqn:radspeed}
\delta v_r \sim P/(\rho H\Omega) \sim c_s,
\end{equation}
where $c_s$ is the fluid sound speed.

For these reasons, in this paper we will be primarily interested in initial warps for which $\phat > 1$ where the disk bends.  Although $H/r$ is not exactly constant throughout our disk, it varies slowly enough that $\phat$ is nearly proportional to $\psi$, and the warp can be characterized in terms of only two parameters, $\phat$ and the radial width of the bend region, $2L_W$.

The parameters for the suite of simulations presented here are given in Table~\ref{tab:simparam}, where the simulation ID encodes the family of simulation, HW and TW for $H/r \approx 1/4$ and $1/10$ respectively, and also gives the approximate value of $\phat$ in the intermediate radial range as well as the radial domain over which the disk is warped.  We use the notation (V)ery (W)ide, (W)ide, and (N)arrow to refer to $L_{W} = 6,4,2$ respectively.  Due to the reduced computational cost of thicker disks, we assembled a suite of HW runs of varying warp amplitudes and radial spans, whereas we limited ourselves to only one TW simulation.  TW is a thinner analogue of $\run{1.5W}$: its warp amplitude and radial span are identical, but the thinner disk results in a larger value of $\phat$.

\begin{figure}
\begin{center}
\includegraphics[width=0.8\textwidth]{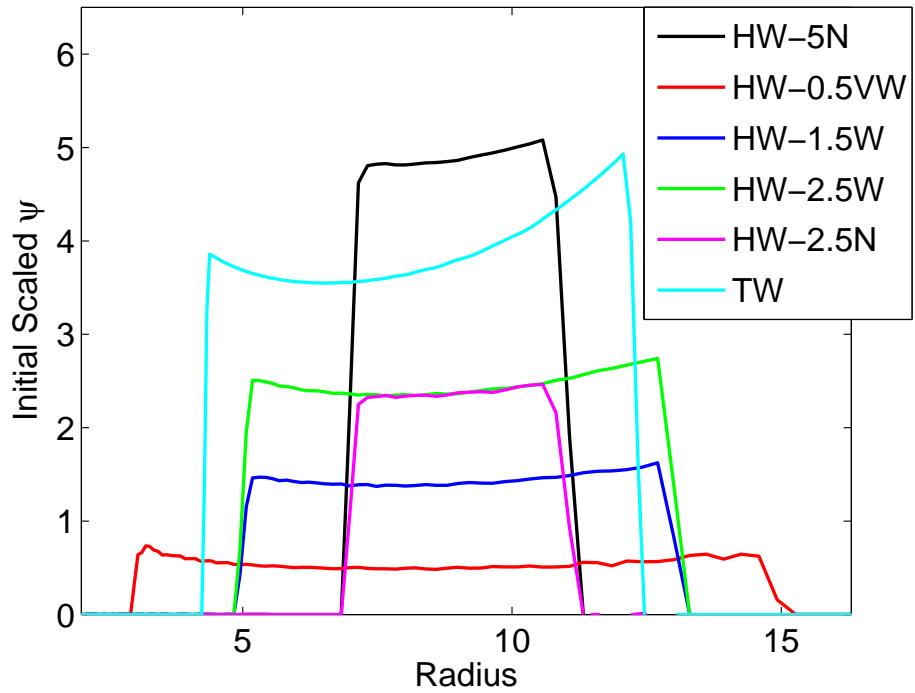}
\caption{Initial radial profile of $\phat(r)$ for all simulations.}
\label{fig:initphat}
\end{center}
\end{figure}

The initial radial profiles of $\phat$ for all the simulations are shown in Figure~\ref{fig:initphat}, where it can be seen that there is little radial variation in $\phat$ in the warping region of the disk.  An example of the warped torus model is presented in Figure~\ref{fig:warpcoord}, in which the density contours of the disk at $\phi = 0$ are shown superimposed on a map of the warped coordinate system ($R',z'$) for simulation $\run{1.5W}$, which we will use as a fiducial model.

\begin{table}
\begin{center}
  \begin{tabular}{@{} |c|c|c|c| @{}}
    \hline
    Run ID & $\theta_{F}$ & $L_{W}$ & Approximate $\hat{\psi}$ \\ 
    \hline
     \run{0.5VW} & $\indeg{12}$ & 6 & 0.5 \\ 
    \run{1.5W} & $\indeg{20}$ & 4 & 1.5 \\ 
    \run{2.5W} & $\indeg{35}$ & 4 & 2.5 \\ 
    \run{2.5N} & $\indeg{15}$ & 2 & 2.5 \\ 
    \run{5N} & $\indeg{35}$ & 2 & 5 \\ 
    \runt & $\indeg{20}$ & 4 & 3.75 \\
    \hline
    
  \end{tabular}
\end{center}
\caption{Simulation parameters.}
\label{tab:simparam}
\end{table}
 
We show an example of the end-result of this procedure in Figure~\ref{fig:warpcoord}, where the initial structure for simulation HW1.5 is shown.

All the simulations were run for 10 orbital periods at the midpoint of the warp, $r_c$, and we use this orbital period as the unit of time.  Thus, in these units the sound speed in the HW simulations is $\approx 14$ and in the TW simulation $\approx 5.6$.  Full 3-D datasets were recorded every 0.1~period.   When we quote averaged values below, we denote a volume-weighted average by $\langle \ldots \rangle$; a density-weighted average is indicated by the notation $\langle \ldots \rangle_\rho$.

\section{Results}
\label{sec:res}


\subsection{Radial Pressure Gradients and Radial Streaming}\label{sec:overview}

\begin{figure}
\begin{center}
\includegraphics[width=0.7\textwidth]{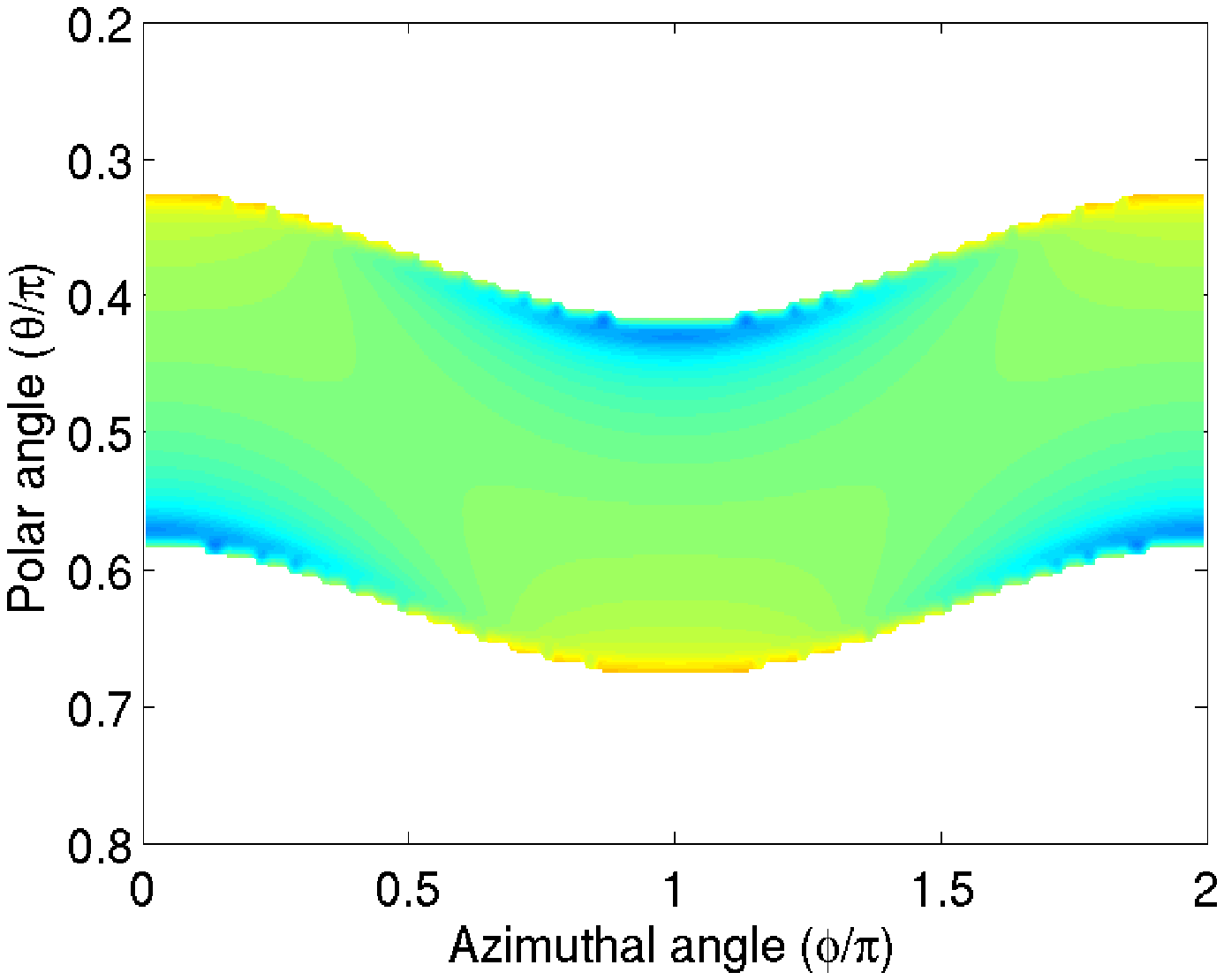} \\
\includegraphics[width=0.7\textwidth]{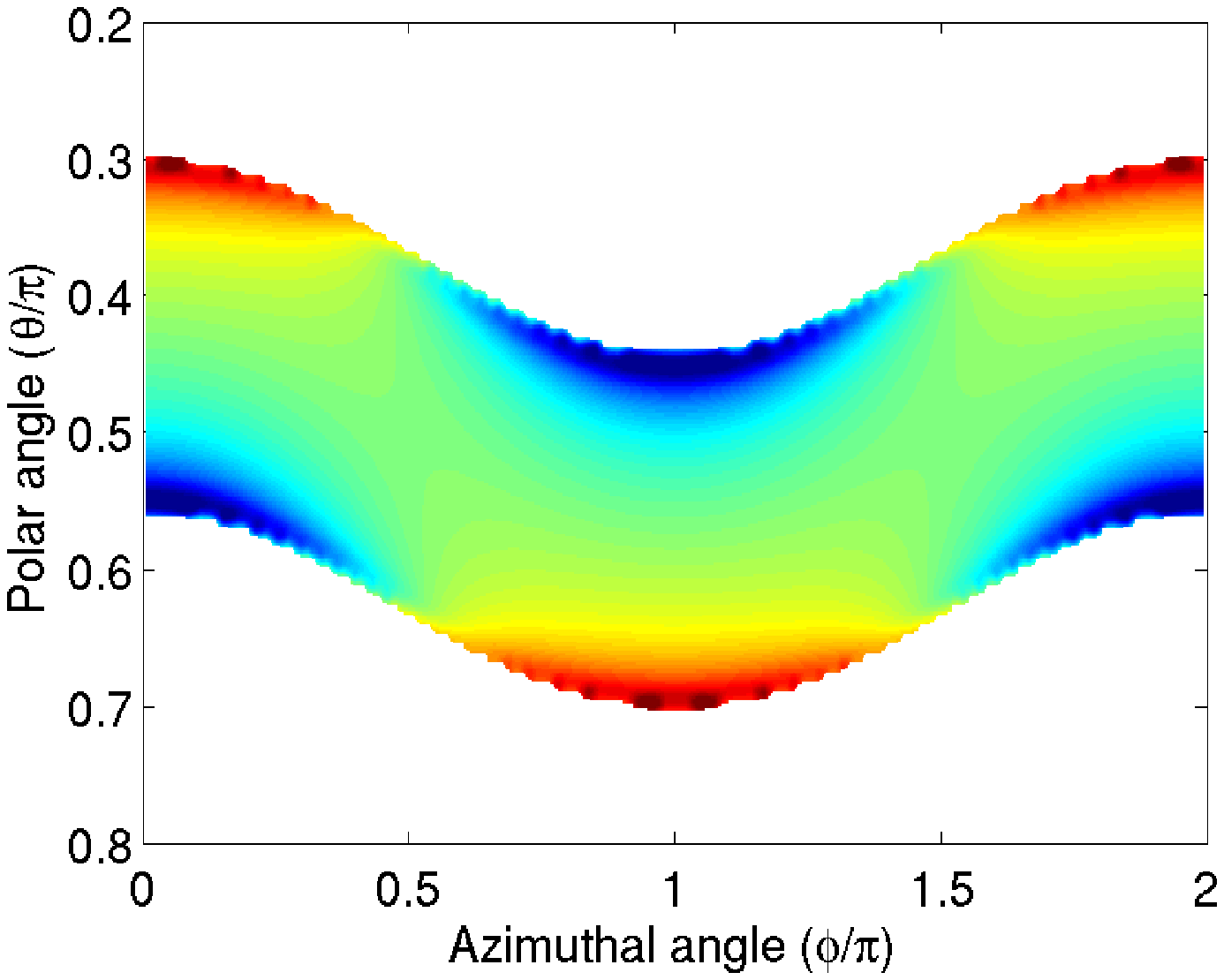} \\
\includegraphics[trim=0cm 0cm 0cm 17cm, clip=true,width=0.9\textwidth]{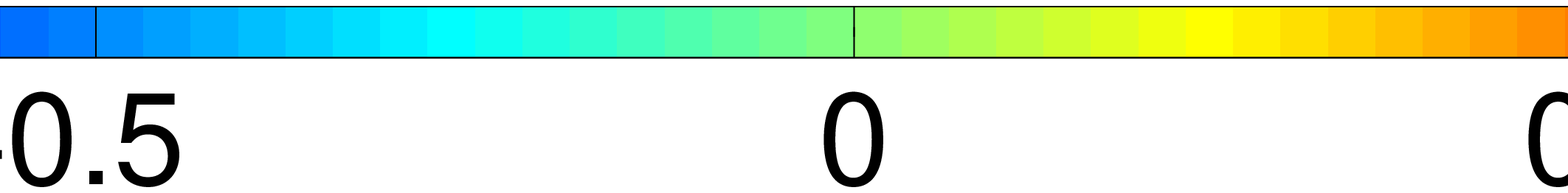}

\caption{Color contours (see color bar) of $d\ln P'/dr$ (Equation~\ref{eqn:deltapgrad}) at $r=9$ at $t=0$ in \run{0.5VW} (Top) and \run{1.5W} (Bottom).  No contours are shown where the density falls below $0.5\%\times$ the maximum.  Coordinates are grid coordinates.}
\label{fig:pressgrad}
\end{center}
\end{figure}

In any hydrostatic flat disk configuration, there are finite radial pressure gradients that balance the gradient in the effective potential due to the slightly non-Keplerian angular momentum distribution.  As expected, in our warped disk there are unbalanced radial pressure gradients right from the start.  Figure~\ref{fig:pressgrad} shows the residual pressure gradient, $d\ln P'/dr$ at radius $r=9$ (the middle of the warp region) at $t=0$ in both the always-linear simulation \run{0.5VW} and the modestly nonlinear simulation \run{1.5W}.  The residual pressure gradient is defined as
\begin{equation}
\frac{d\ln P'}{dr} = \frac{d\ln P}{dr} -  \frac{d\ln P_0}{dr},
\label{eqn:deltapgrad}
\end{equation} 
where $d\ln P_0/dr$ is the midplane value of $d\ln P/dr$ in an unwarped torus.  Within 1--2 scale heights of the midplane, $d\ln P_0/dr$ changes minimally, so it is also a very good approximation to the radial pressure gradient even well off the midplane.  At $r=9$, we find $d\ln P_0/dr \approx -0.4$. We make this adjustment to the pressure gradient in order to emphasize the portion of it due to the warp.  

In the linear simulation, we see that the warp induces only small deviations about $d\ln P_0/dr$.  In the weakly nonlinear simulation, these deviations are greater and are asymmetric about the midplane.  This asymmetry is a result of the asymmetric geometry of nonlinear warping.  In the $\phi=0$ slice shown in Figure~\ref{fig:warpcoord}, the ``top'' of the disk has an outward radial pressure gradient (and therefore an inward pressure force), whereas the ``bottom'' of the disk has an outward pressure force.  Near the disk surface in the warped region, the pressure gradients are always larger on the top.  These relations are, of course, reversed at $\phi = \pi$.  As expected, the gradient is weak everywhere in the linear case.  On the other hand, the gradient in parts of the nonlinear simulation is sharp enough that its scale length is $\sim 1$ in our distance units, $\sim r/9 \sim H/2$ at the radius shown.  Thus, the radial pressure gradient is large enough to drive transonic radial motions when $\hat\psi > 1$.    



\begin{figure}
\begin{center}
\includegraphics[width=0.7\textwidth]{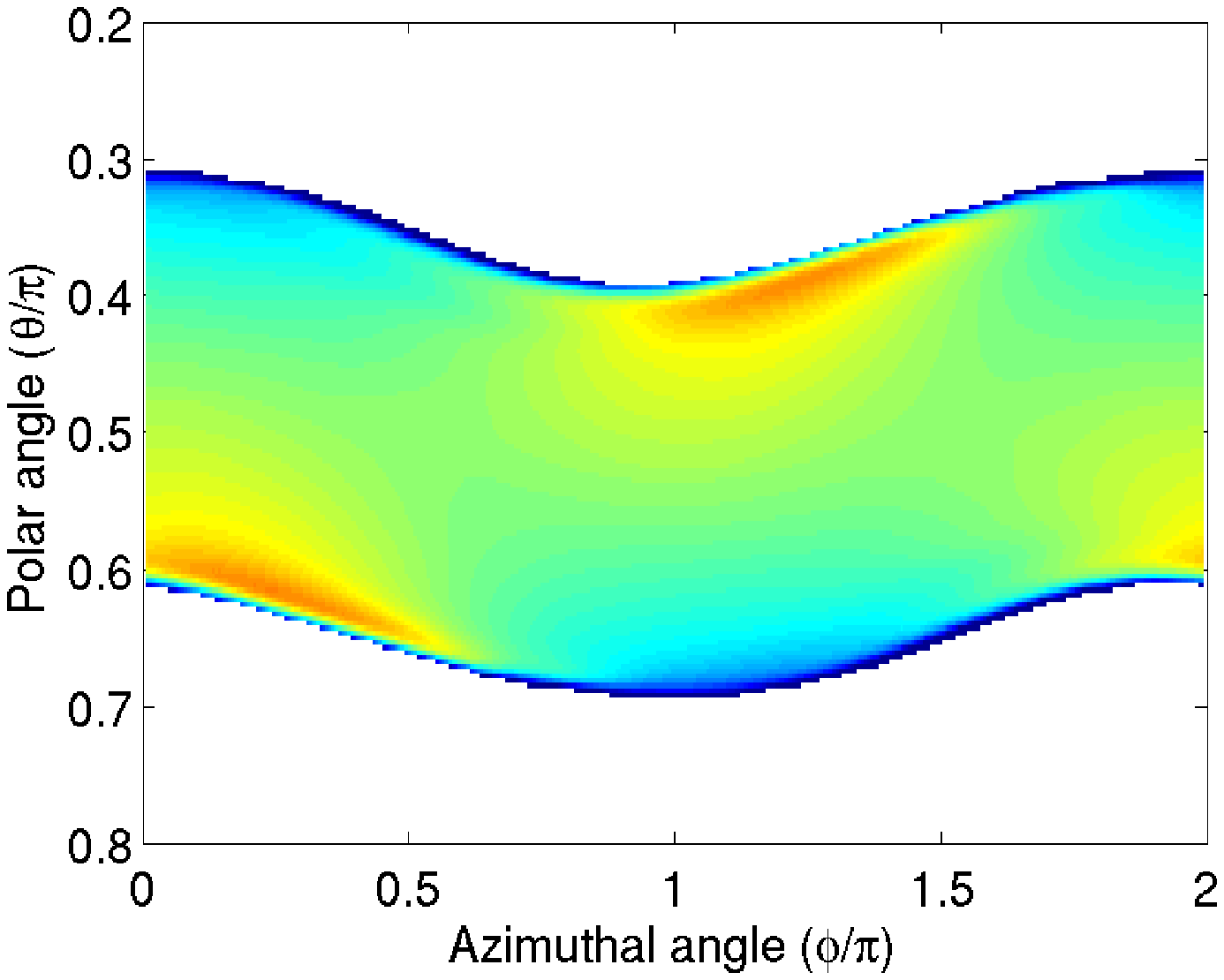} \\
\includegraphics[width=0.7\textwidth]{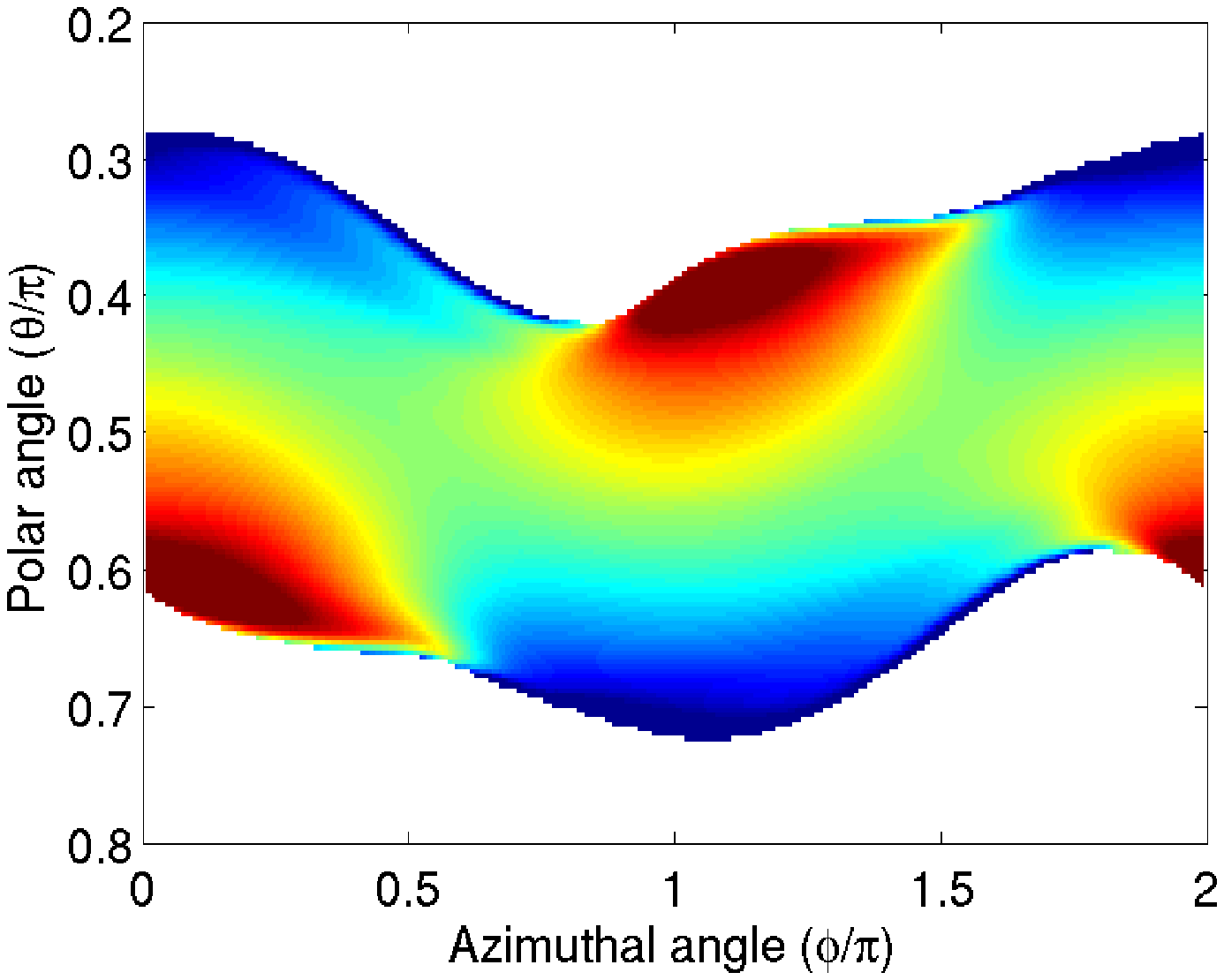} \\
\includegraphics[trim=0cm 0cm 0cm 17cm, clip=true,width=0.9\textwidth]{fig3c.eps}
\caption{Color contours (see color bar) of $v_r/c_s$ at $r=9$ at $t=0.5$ in \run{0.5VW} (Top) and \run{1.5W} (Bottom).  No contours are shown where the density falls below $0.5\% \times$ the maximum.  Coordinates are grid coordinates.}
\label{fig:transonic}
\end{center}
\end{figure}

That these pressure gradients are effective in driving transonic motions is demonstrated by the data shown in Figure~\ref{fig:transonic}.  This figure portrays the situation at the same locations shown in Figure~\ref{fig:pressgrad} half an orbit after the beginning of the simulation.  In this time, fluid elements have traversed a range of $\pi$ in azimuthal angle, sampling the full range of variation of the radial pressure gradient shown in Figure~\ref{fig:pressgrad}.  In the linear case, there are subsonic fluid motions symmetric about zero.
In the weakly nonlinear simulation, we see that throughout a significant fraction of the disk volume (and a slightly smaller fraction of its mass), the initial radial pressure gradients have induced near-sonic radial motions consistent with the orientation of the initial pressure gradient.  The magnitude of inward and outward radial motion differs due to the asymmetry of the initial radial pressure gradient induced by the warp.


\begin{figure}
\begin{center}
\includegraphics[width=0.5\textwidth]{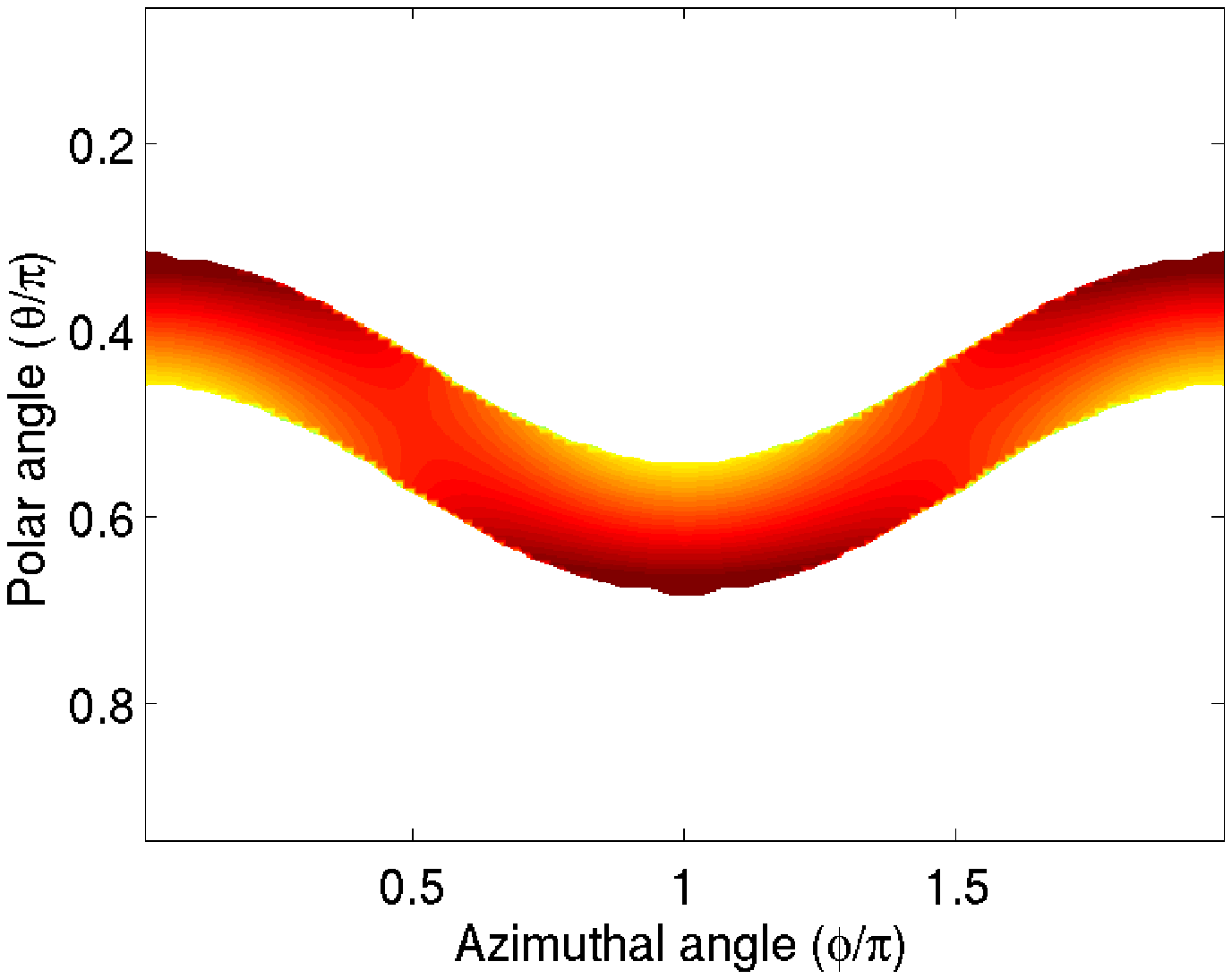} \\
\includegraphics[width=0.5\textwidth]{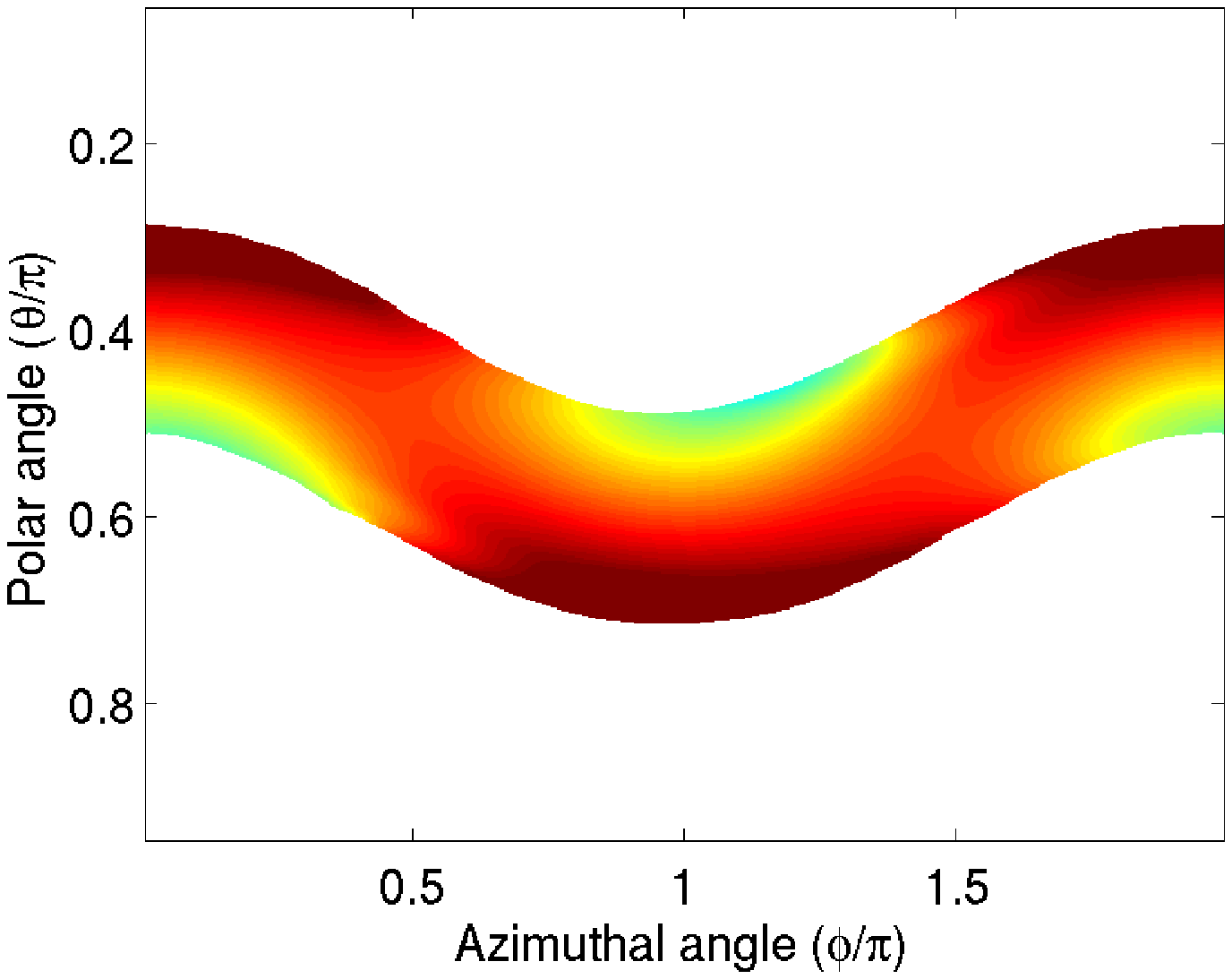} \\ 
\newpage
\includegraphics[width=0.5\textwidth]{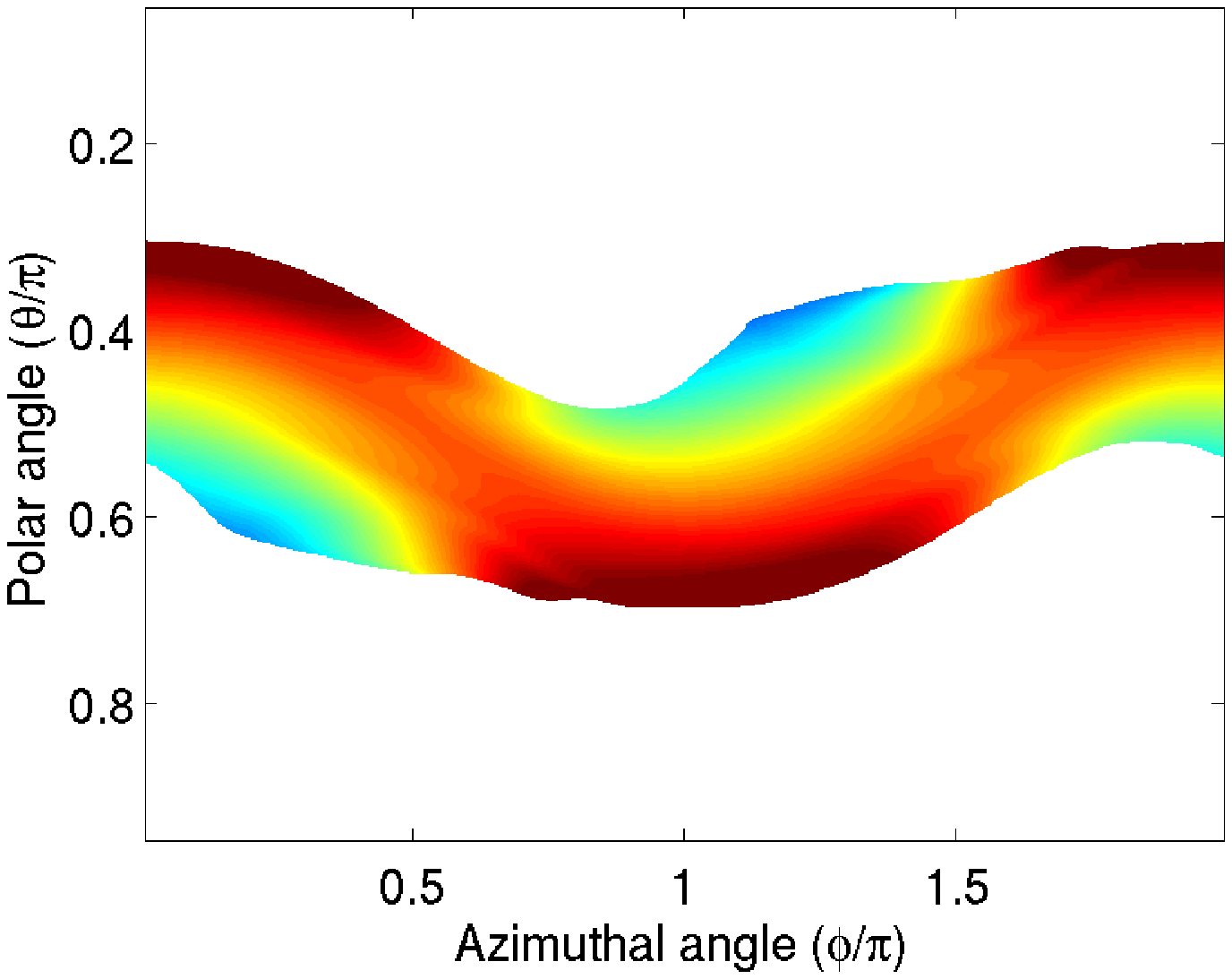} \\
\includegraphics[trim=1cm 1cm 1cm 24cm, clip=true,width=0.6\textwidth]{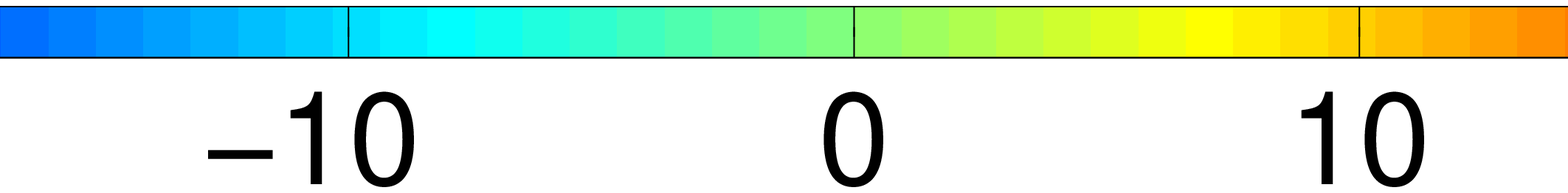}

\caption{Color contours of $\tan^{-1}(-L_x/L_z)$ on the shell at $r=14$ from simulation \run{1.5W}.  Three times are shown: the initial condition, 0.5 orbits, and 1 orbit.  Only cells where the local density is greater than $0.5\%$ of the maximum are illustrated.  Coordinates are grid coordinates.}
\label{fig:mix}
\end{center}
\end{figure}

  It is these rapid radial motions that mix fluid elements with differing initial orientation of angular momentum, ultimately causing the entire disk to align in a single plane.  The progress of this mixing can be seen in the three panels of Figure~\ref{fig:mix}.
There is always a vertical gradient in the orientation of the angular momentum due simply to geometry: even when all the orbital velocities are parallel, the radius vector from the origin swings with altitude across the disk.  The magnitude of this purely geometric effect can be seen by the range of orientation vertically through the disk at a fixed azimuthal angle $\phi$.  However, as the disk evolves, the mean orientation at a given $\phi$ changes and the range of orientations increases.  Because there are no external torques on this system, the only way the fluid at a fixed location can change the orientation of its angular momentum is by transport of material with different angular momentum to that location.  This evolution is therefore the signature of angular momentum mixing.  Moreover, because the range of orientations seen at 1 orbit includes directions not present at all at that radius in the initial state, there must be radial fluid motions to convey the matter with the new orientation.    

\begin{figure}
\begin{center}
\includegraphics[width=0.8\textwidth]{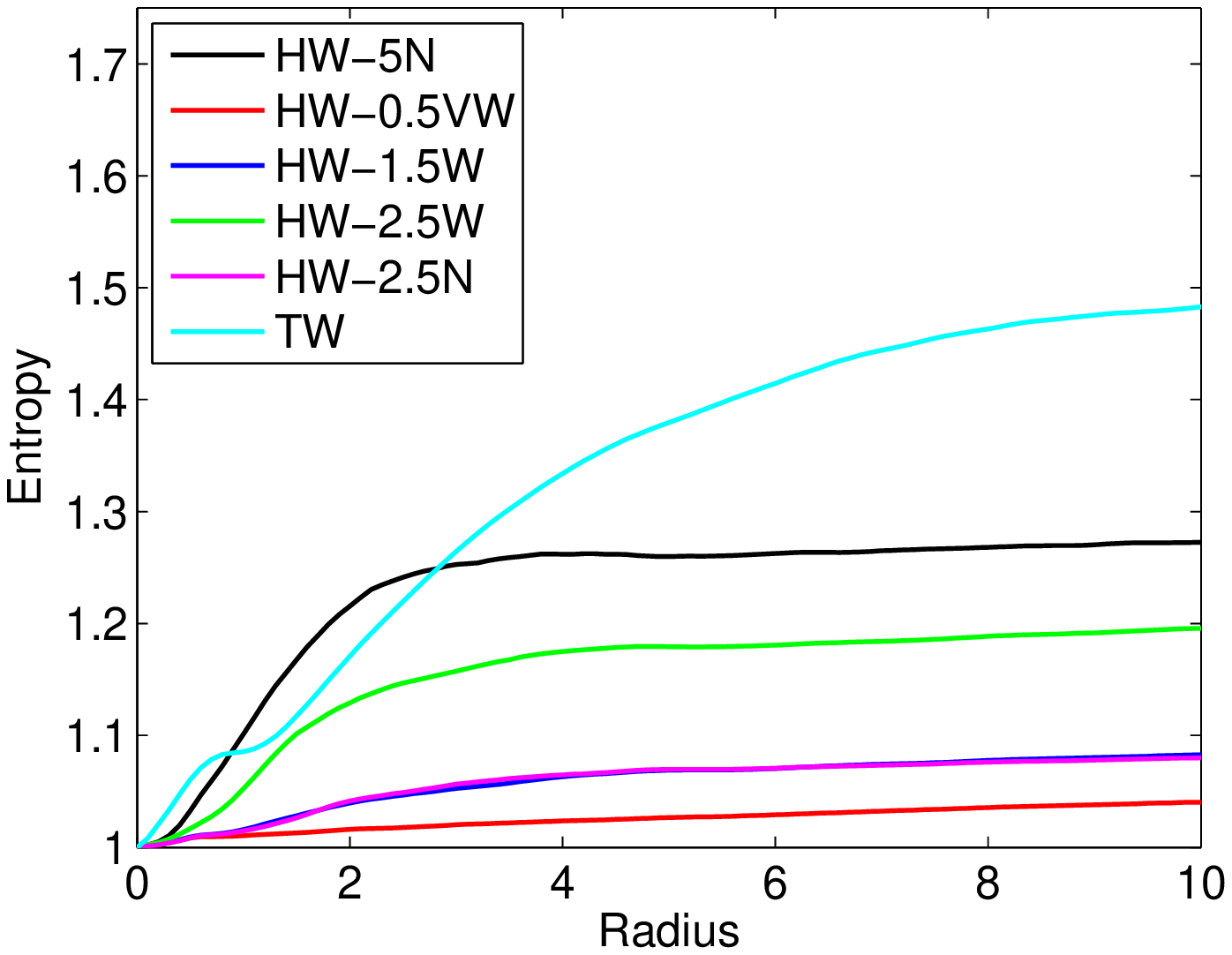}
\caption{Evolution of volume-integrated entropy for all simulations scaled to its initial value.}
\label{fig:entevol}
\end{center}
\end{figure}

As we have shown, radial pressure gradients associated with disk warping drive transonic radial motions; when these streams intersect, their angular momentum is mixed and the warp is dissipated.  In the traditional approach to this topic, the speed of radial motions is limited primarily by a phenomenological isotropic velocity parameterized by $\alpha$; here, the ultimate speed of radial motions is limited by a combination of achieving the maximum speed possible for acceleration by a thermal pressure gradient (i.e., Mach number order unity) and the bulk viscosity that appears only in regions of strong compression (i.e., shocks).  Figure~\ref{fig:entevol} displays the rate at which these shocks create entropy.  Because strongly nonlinear behavior persists for only a small number of orbits in the bulk of the disk, essentially all the entropy production is accomplished in the first two orbits.  Contrasting different simulations, we find that the absolute amount of entropy production rises with both increasing $\meanphat$ and radial extent of the warp.

\subsection{Rate of Relaxation Toward the Mean Plane}\label{sec:quant}

The end-result of the relaxation process must be alignment in a single plane because, given the constraint of angular momentum conservation and the lack of any external torque, that is the only possible long-term equilibrium.  Although it is true that in the limit of very small amplitude, linear bending waves can propagate indefinitely if there is no dissipative process, nonlinear waves---like the ones studied here---create pressure gradients that cause momentum exchange and, in effect, mix angular momentum.  The tilt of the resulting orbital plane relative to the equatorial plane must be $\theta_{T}(r) = \theta_{M}$, where
\begin{equation}\label{eqn:theta_m}
\theta_{M} = \tan^{-1} \left( \myfrac{< - L_{x}>}{<L_{z}>} \right).
\end{equation}
Here the local angular momentum is $\vec L$ and the components used in equation~\ref{eqn:theta_m} are in terms of the grid coordinates.
To quantify the general level of deviation from this expected equilibrium, we define two measures.  One is $ \delta \theta_{T} (r,t)$, 
\begin{equation}
\delta \theta_{T} (r,t) = \theta_{T}(r,t) - \theta_{M},
\end{equation}
the local angular deviation from the equilibrium state.  The other is a mass-weighted spatial average of $\hat\psi$, the normalized radial derivative of the orientation angle
\begin{equation}
\meanphat (t) = \myfrac{<\rho \phat(r,t)>}{<\rho>}.
\end{equation}

\begin{figure}
\begin{center}
\includegraphics[width=0.8\textwidth]{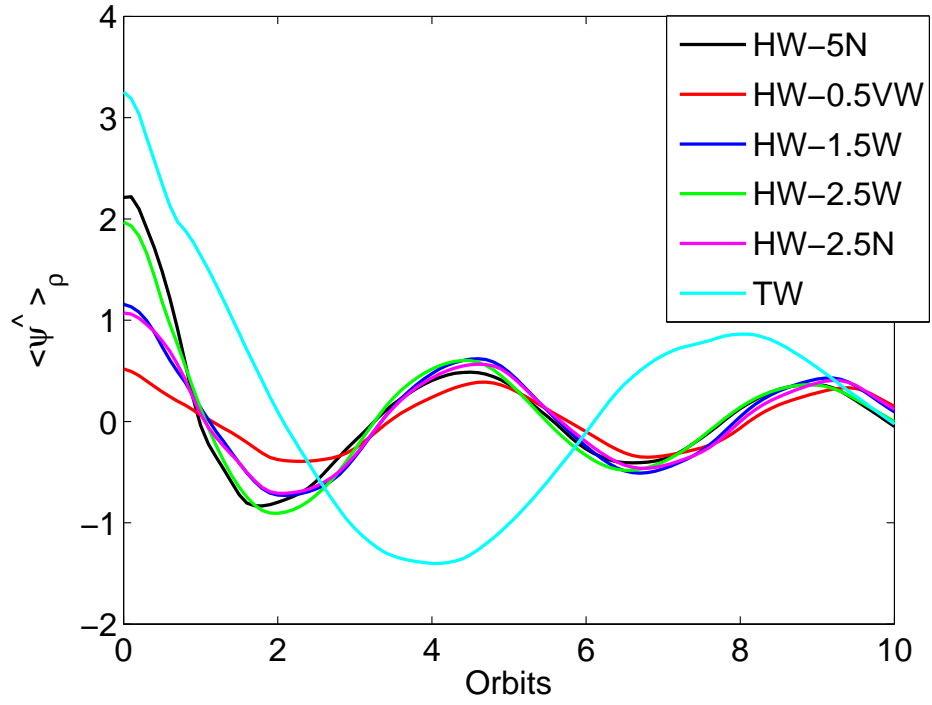}
\caption{Evolution of $\langle\phat\rangle_\rho$ for all simulations.}
\label{fig:dtt}
\end{center}
\end{figure}

The time evolution for the latter quantity is displayed in Figure~\ref{fig:dtt}.  Several clear patterns can be seen in this figure.  The first is that all the runs exhibit an oscillation in $\meanphat$.  In all those runs whose initial state was nonlinear, the amplitude of this oscillation decays rapidly at first, but once $\meanphat$ drops below $\simeq 1$, it varies much more slowly for the remainder of the simulation; the initially linear simulation oscillates with nearly constant amplitude from the start.  In other words, $\phat > 1$ is not only an indicator that transonic radial motions will be created, it is also a semi-quantitative predictor of the rate at which the warp amplitude relaxes: this rate becomes much slower when $\phat$ falls below 1.

The oscillation is a bending wave induced by the initial warp; in fact, one way of describing our results is that an initially nonlinear bending wave rapidly decays to linear amplitude.  The oppositely directed radial motions created by the warp on the top and bottom edges of the disk set up the circulatory flow that characterizes these waves.  The mode induced is the fundamental (i.e., the radial extent of the disk is half a wavelength) because the disk inclination relative to its mean value varies monotonically from one radial extreme of the disk to the other. 

It is instructive to model the time-dependent behavior of this oscillation in terms of a function $f(t)$ of the form
\begin{equation}
f(t) = A_{0}\cos(\omega t)\exp(-s t).
\label{eqn:nlmod}
\end{equation}
The fitting itself can be done in a variety of ways, however we find that the fit parameters ($A_{0},\omega,s$) depend only weakly on the choice of method.  Our method determines the frequency of the oscillation $\omega$ using the second zero of $\meanphat$; we find the parameter $s$ from the magnitude of attenuation at the first minimum of $\meanphat$; $A_{0}$ is simply determined by the initial value of $\meanphat$.  The fit parameters are shown in Table~\ref{tab:nonlin}, with $\omega$ being replaced by the corresponding period, $P=2\pi/\omega$, and $s$ in units of inverse orbital periods.

\begin{table}
\begin{center}
  \begin{tabular}{@{} |c|c|c|c|c| @{}}
    \hline
    Run ID & $A_0$ & P & $s$  \\ 
    \hline
    \run{0.5VW} &  $0.52$ & $4.60$ &  $0.12$ \\ 
    \run{1.5W}  &  $1.16$ & $4.47$ &  $0.22$ \\ 
    \run{2.5W}  &  $1.97$ & $4.33$ &  $0.39$ \\ 
    \run{2.5N}  &  $1.07$ & $4.47$ &  $0.20$ \\ 
    \run{5N}    &  $2.21$ & $4.33$ & $0.54$ \\ 
    \runt       &  $3.25$ & $8.20$ & $0.21$ \\
    \hline
    
  \end{tabular}
\end{center}
\caption{Quantifying nonlinear warp relaxation.}
\label{tab:nonlin}
\end{table}

The most prominent result of this fitting exercise is that over the entire parameter space we have studied, relaxation in the nonlinear phase is quite rapid, requiring only a few orbits at the mid-point of the disk (the characteristic decay times $s^{-1}$ are $\simeq 2$--5~orbits, rather longer than the duration of the relaxation, because the initial values of disk-integrated $\langle \phat \rangle_\rho$ are relatively small, $\simeq $1--2, and the rapid decay persists only until $\langle \phat \rangle_\rho \lesssim 1$).  The rate of relaxation does, however, vary from case to case.  If the ``severity" of the bend is some compound of $\phat$ and the radial width over which the disk bends so sharply, the nonlinear relaxation rate correlates with ``severity" in the way one might expect: more dramatic bends relax more rapidly.  At fixed $L_W$, the fractional decay rate $s$ scales with $\psi$ roughly $\propto \psi^{b}$, with $b \simeq 1$--1.5.  When $L_W$ changes, a larger span of warping leads to a larger decay rate; for example, the decay rate of \run{2.5W} is almost exactly twice the rate exhibited by \run{2.5N}, which shares its $\meanphat$ but whose warp stretches only half as far.  Conversely, \run{5N} and \run{2.5W} have similar relaxation rates even though their values of $\meanphat$ are a factor of 2 different; they decay similarly because the case with smaller $\meanphat$ has a warp that is twice as wide. A similar pairing exists between \run{2.5N} and \run{1.5W}.  These pairings are also evident in the rate of entropy production (see Fig.~\ref{fig:entevol}).

However, comparing the relaxation rates for \run{1.5W} and \runt\ reveals something new and unexpected.  Following the pattern in which the decay rate scales with $\phat$, the rate at which the amplitude diminishes in \runt\ is considerably faster than in \run{1.5W}.  However, their {\it fractional} decay rates $s$ are nearly identical.  These two simulations began with identical bending rates $\psi$, differing only in sound speed.  Therefore, in this regime, the fractional decay rate $s$ depends on $\psi$, not $\phat$ or the sound speed.  This is surprising because the primary driver of radial motions is unbalanced pressure gradients, so one might have thought that the speed at which fluid elements could mix their angular momenta would always scale with the sound speed, yet here the mixing rate appears to have no dependence on the sound speed at all.

That this is possible may have to do with the specifics of our simulations.  Typical Mach numbers in \run{1.5W} during the nonlinear relaxation period are only $\simeq 0.5$--1; at the equivalent time in \runt, they are $\simeq 1$--3.  Both are order unity, but the factor of $\sim 3$ between the characteristic Mach numbers compensates for the factor $\sim 3$ in the opposite direction between their sound speeds.   Perhaps if the sound speed were reduced another factor of 3 (and therefore would require rather greater grid-cell resolution than we can reach at the moment), the Mach numbers achieved during the nonlinear relaxation period would not be terribly much greater than in \runt \ because it is very difficult for pressure forces alone to accelerate motions to more highly supersonic speeds.  If so, when $\phat \gtrsim 5$--10, the mixing speeds may reach a limiting value of several times the sound speed.  At this point, all we can say is that within the range of parameters spanned by our simulations, $s \simeq 0.14 (\psi/0.6)^{b}(L_W/r)\Omega$ with $b \simeq 1$--1.5, with no dependence on $c_s$.

\begin{figure}
\begin{center}
\includegraphics[width=0.8\textwidth]{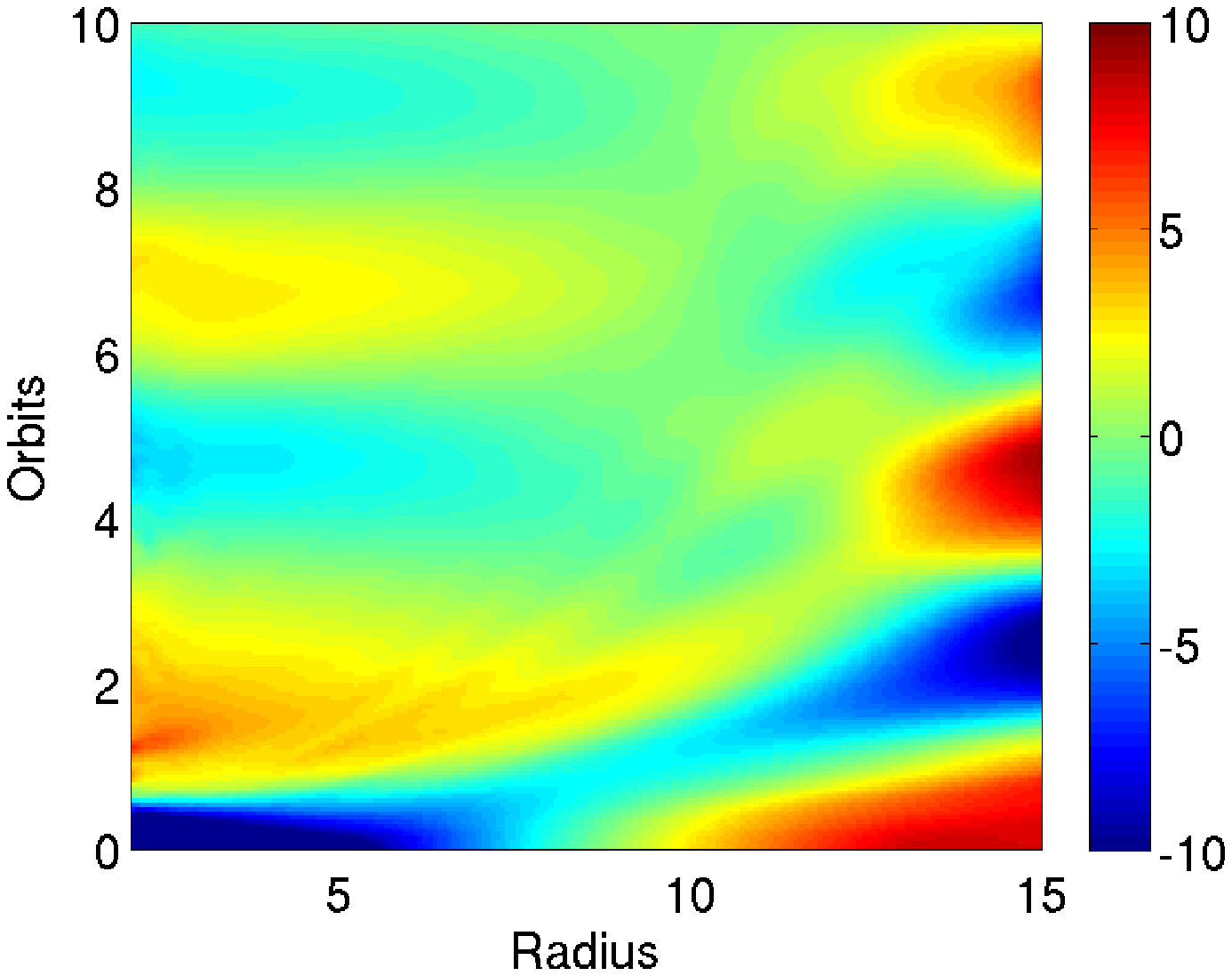} \\
\includegraphics[width=0.8\textwidth]{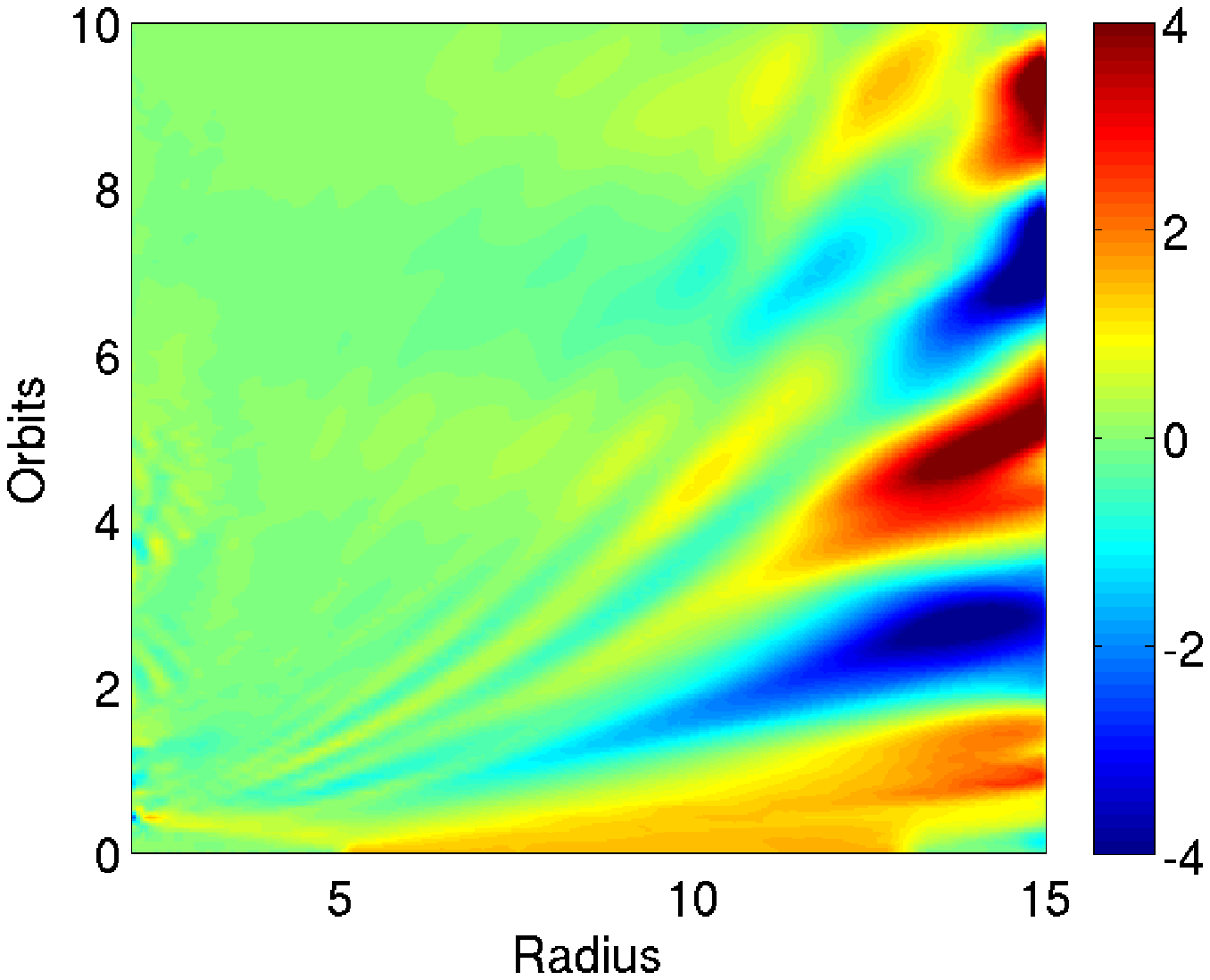}
\caption{Spacetime diagram of $\dtt(r,t)$ (Top, measured in degrees) and $\hat{\psi}(r,t)$ (Bottom) from simulation $\run{1.5W}$.}
\label{fig:dttst}
\end{center}
\end{figure}

Internal details of the relaxation and bending wave phases can be seen in Figure~\ref{fig:dttst}, which displays $\dtt(r,t)$ and $\hat{\psi}(r,t)$ from simulation $\run{1.5W}$.  The quantity $\dtt$ shows the shape of the bending wave more clearly; $\phat$ quantifies its degree of nonlinearity.  In its initial state, the disk bends down at small radii and up at large, reflecting its transition in orientation angle, as seen best in $\dtt$.  The local normalized bending rate $\phat$, on the other hand, is initially zero at both large and small radii and comparatively large in the middle.  Because this simulation was only weakly nonlinear even in its initial state, $\phat$ in most of the disk quickly becomes less than unity.  As the disk relaxes from its nonlinear warp, it falls into a bending wave pattern corresponding to the normal mode whose wavelength is twice the radial extent of the disk because this is the mode most closely related to the initial state (once again, seen best in $\dtt$).  Radially-dependent Fourier analysis demonstrates that these oscillations have the same frequency at all radii, confirming that these oscillations constitute a normal mode of the disk.  Even during the later, predominantly linear phase, both $\dtt$ and $\phat$ remain large at large radii because the disk's inertia decreases sharply toward the edges;
regions of low inertia have high amplitude because the wave action is conserved along the direction of propagation.  Wave action conservation alone would lead to nonlinear amplitude at both edges, but the nonlinear damping we have already discussed acts on the orbital timescale, and is therefore much more rapid at $r=2$ than at $r=15$.  As a result, the wave amplitude is relatively small at the inner edge even while it remains nonlinear at the outer edge.

\subsection{The Angular Momentum Flux}

In order for the disk to arrive at a flat configuration, it must mix angular momentum from the different portions of the disk so that all regions have angular momentum with the same direction.  To describe most clearly how this occurs, it is convenient to define a pair of coordinate systems, one spherical and the other Cartesian, both oriented according to the mean angular momentum $\langle \vec L \rangle$ of the disk.  To distinguish this Cartesian system from the one associated with the grid coordinates, we designate the axes of this one by upper-case letters.  We will designate the direction of $\langle \vec L \rangle$ as the polar axis of the spherical system and the $Z$-axis of the Cartesian system.  To define the zero-point of azimuthal angle for the spherical system and the $X$-axis of the Cartesian system, we choose a direction that keeps the tilt in the $X$-$Z$ plane.  Thus, the dynamics of relaxation can be thought of as the redistribution of $L_X$ of differing signs until the local $L_X = 0$ everywhere.

Ordinarily, hydrodynamic momentum fluxes are discussed in terms of a Reynolds stress tensor ${\bf R} = \rho \vec v \otimes \vec v$.  When described in coordinate language, the two velocity vectors are nearly always projected onto the same basis.  Here, however, we are most concerned with the flux of angular, rather than linear, momentum, and the radial motion of the Cartesian component $L_X$.  We therefore define the angular momentum flux tensor
\begin{equation}
{\bf S} \equiv \int \, dA \, \rho \vec v \otimes \vec L,
\end{equation}
describing $\vec v$ in terms of spherical coordinates, but $\vec L$ in terms of the Cartesian coordinates just defined.   The integral is taken over a spherical surface at radius $r$.  In this language, the element of greatest interest in this tensor is $S_{rX}$.  A natural unit for this tensor is given by $S_0(r) = \int \, dA \, r P$.  We can study the dependence of $S_{rX}/S_0$ on parameters by looking at variations relative to a fiducial case, which we choose to be \run{2.5W}.

\begin{figure}
\begin{center}
\includegraphics[width=0.4\textwidth,angle=90]{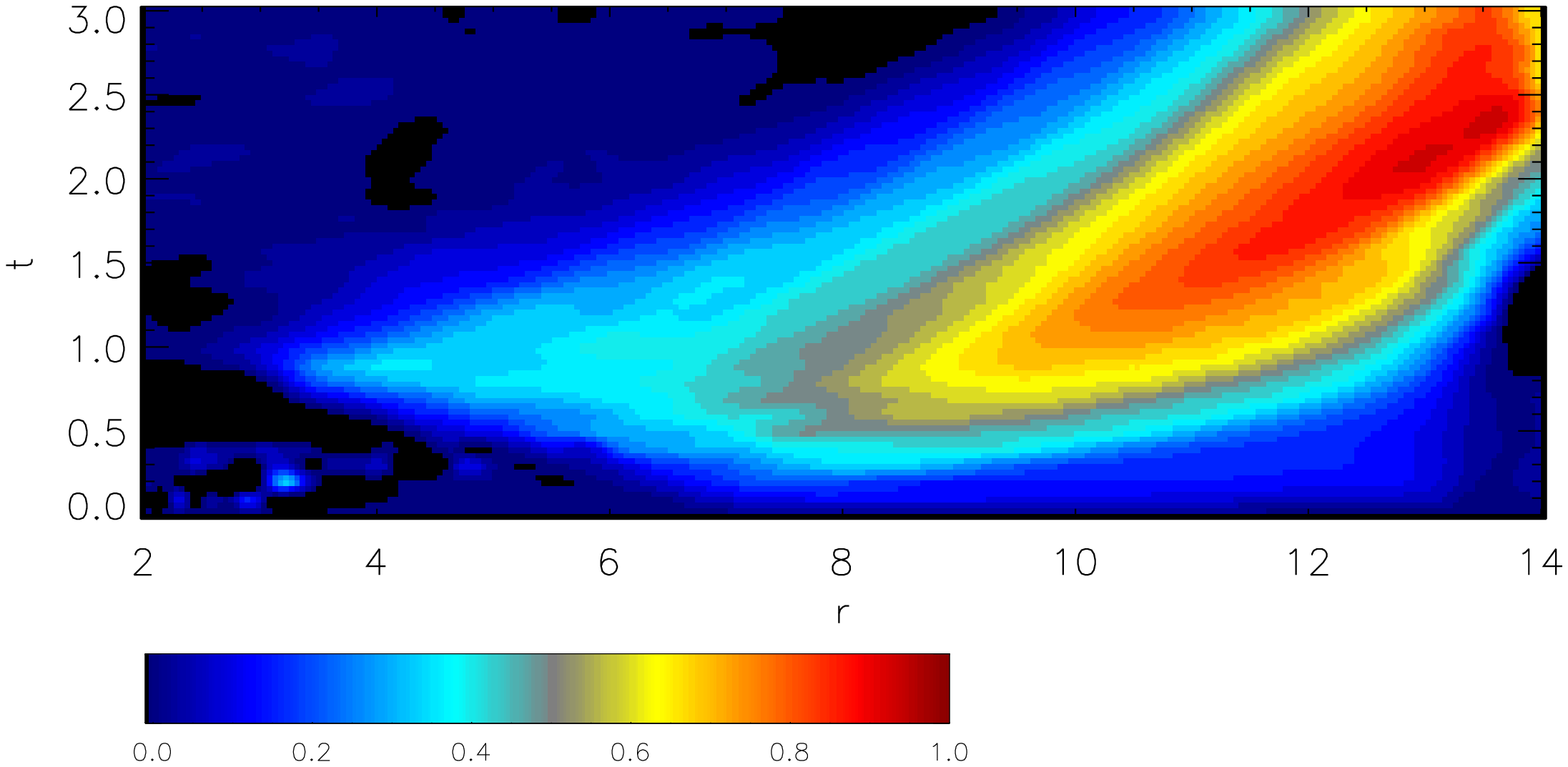}
\includegraphics[width=0.4\textwidth,angle=90]{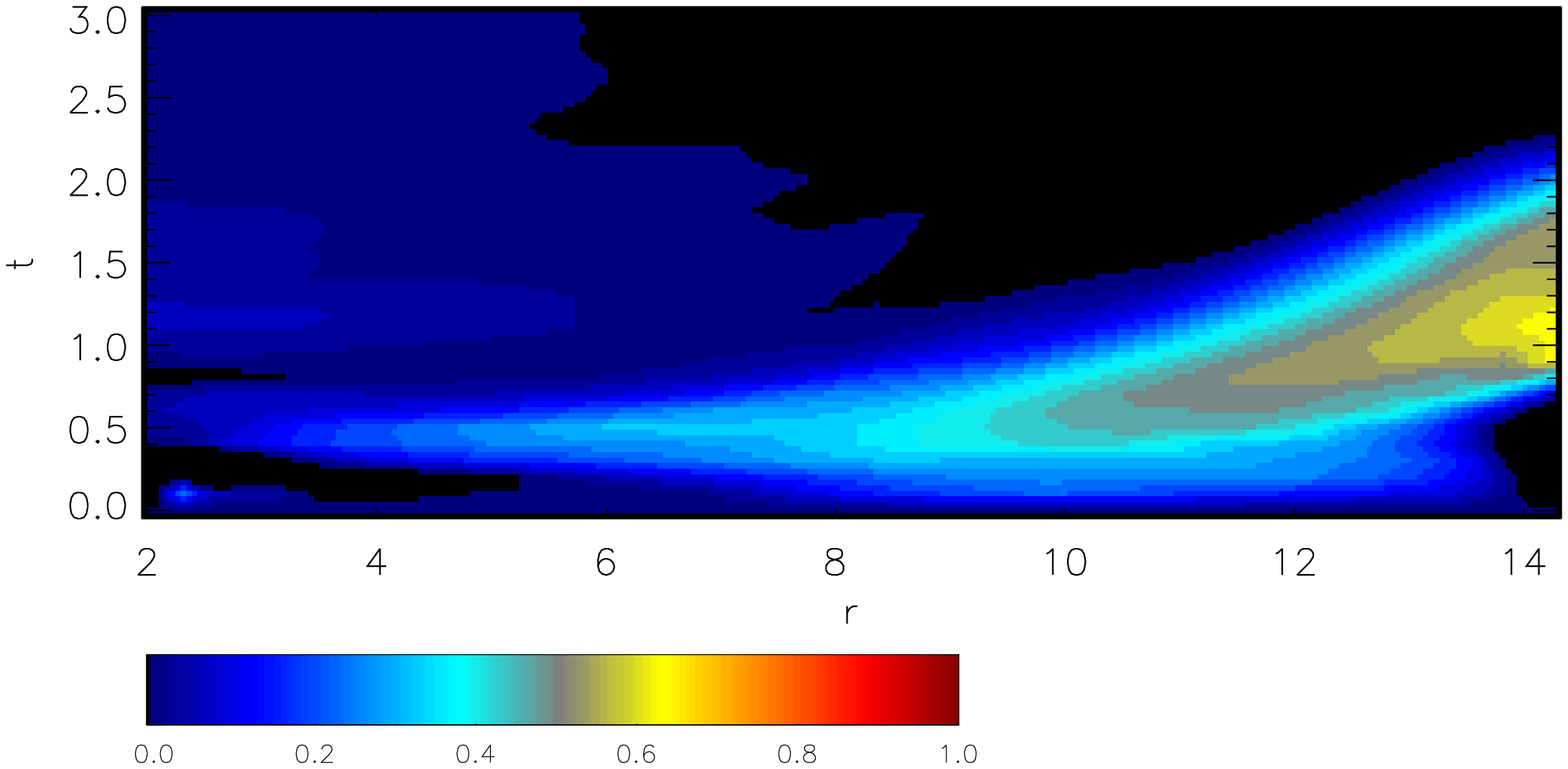}
\includegraphics[width=0.4\textwidth,angle=90]{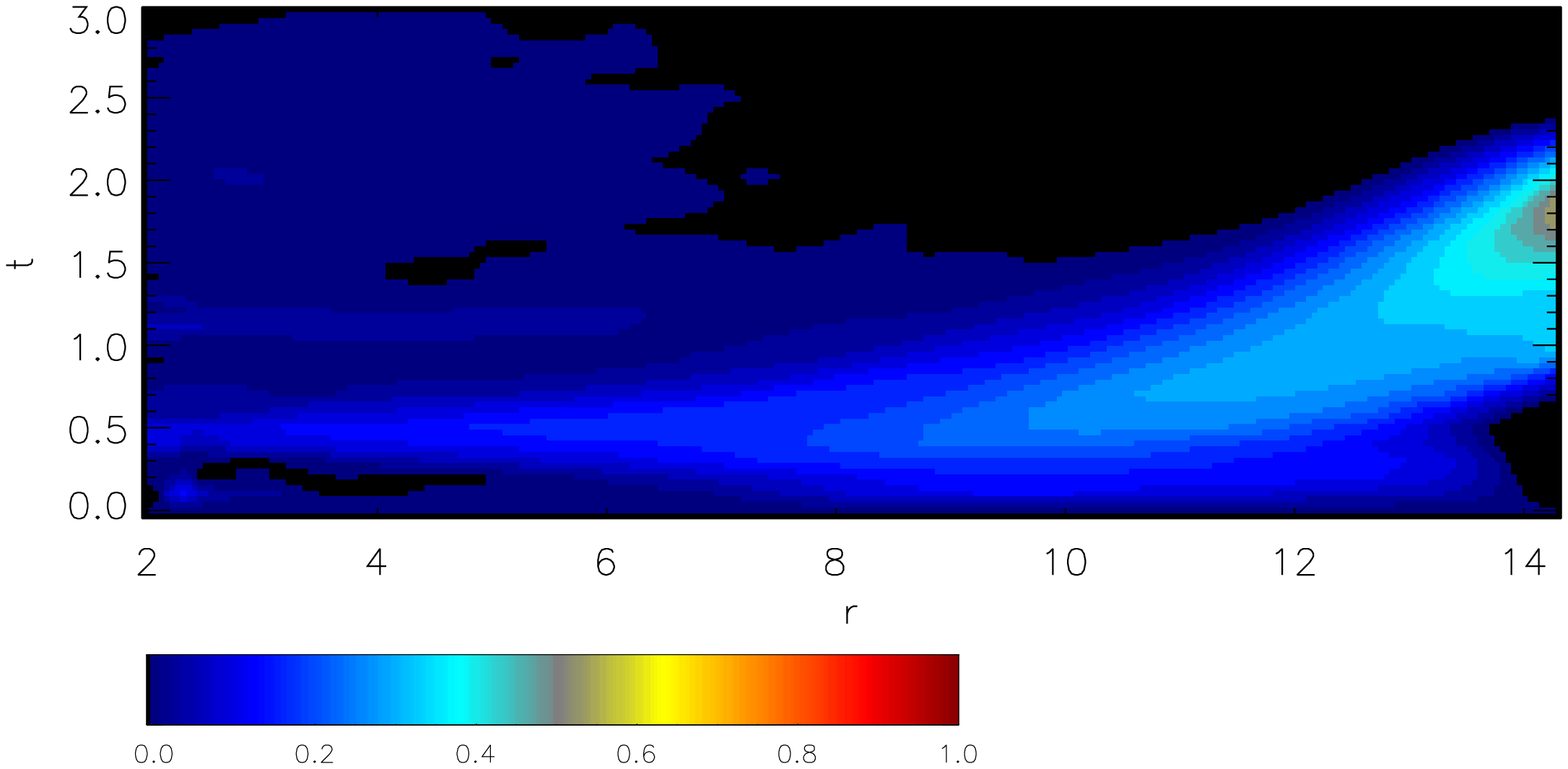}
\caption{The evolution in $r$ and $t$ of the normalized angular momentum flux $S_{rX}/S_0$.  Three simulations with the same $L_W$ are shown: \runt\ (top panel), \run{2.5W}
(middle panel) and \run{1.5W} (bottom panel).}
\label{fig:Srx_equalW}
\end{center}
\end{figure}

Figure~\ref{fig:Srx_equalW} shows how $S_{rX}/S_0$ evolves as a function of $r$ and $t$ during the principal relaxation stage in three of our simulations that all have the same warp width: \runt, \run{2.5W}, and \run{1.5W}.  As this figure illustrates, $S_{rX}/S_0$ is $>0$ at all radii for virtually all of the relaxation process, as positive $L_X$ is taken outward from small radii and negative $L_X$ is taken inward from large radii.  In magnitude, it reaches peak values of $\simeq 0.5$--1, depending on the parameters of the simulation.  However, these peak values are generally found at the outer edge of the disk, not the center of the warp.  Near $r = r_c$, the peak value of $S_{rX}/S_0$ ranges from $\simeq 0.15$--0.65.  It is also clear from this figure that the relaxation takes only a brief time, never more than 1--3 orbits at the central radius of the warp.

The data of Figure~\ref{fig:Srx_equalW} can also be used to uncover how the peak magnitude of the stress at the center of the warp depends on parameters.  Comparing \run{1.5W} and \run{2.5W} shows that, in rough terms, $S_{rX}/S_0$ is proportional to the initial warp strength $\phat$. However, the fact that the normalized peak stress at $r_c$ increases by a factor $\simeq 4$ from \run{1.5W} to \runt, which share the same warp geometry factor $\psi = d\theta/d\ln r$, shows that the peak magnitude of the normalized stress is even more strongly dependent upon the sound speed, increasing at a rate between linear and quadratic as the sound speed decreases.   Put another way, the peak {\it unnormalized} stress decreases, but only weakly, with decreasing sound speed.  That this should be so is another reflection of the fact already pointed out in the previous subsection, that the Mach number of the radial motions can increase sharply when the sound speed decreases.

\begin{figure}
\begin{center}
\includegraphics[width=0.4\textwidth,angle=90]{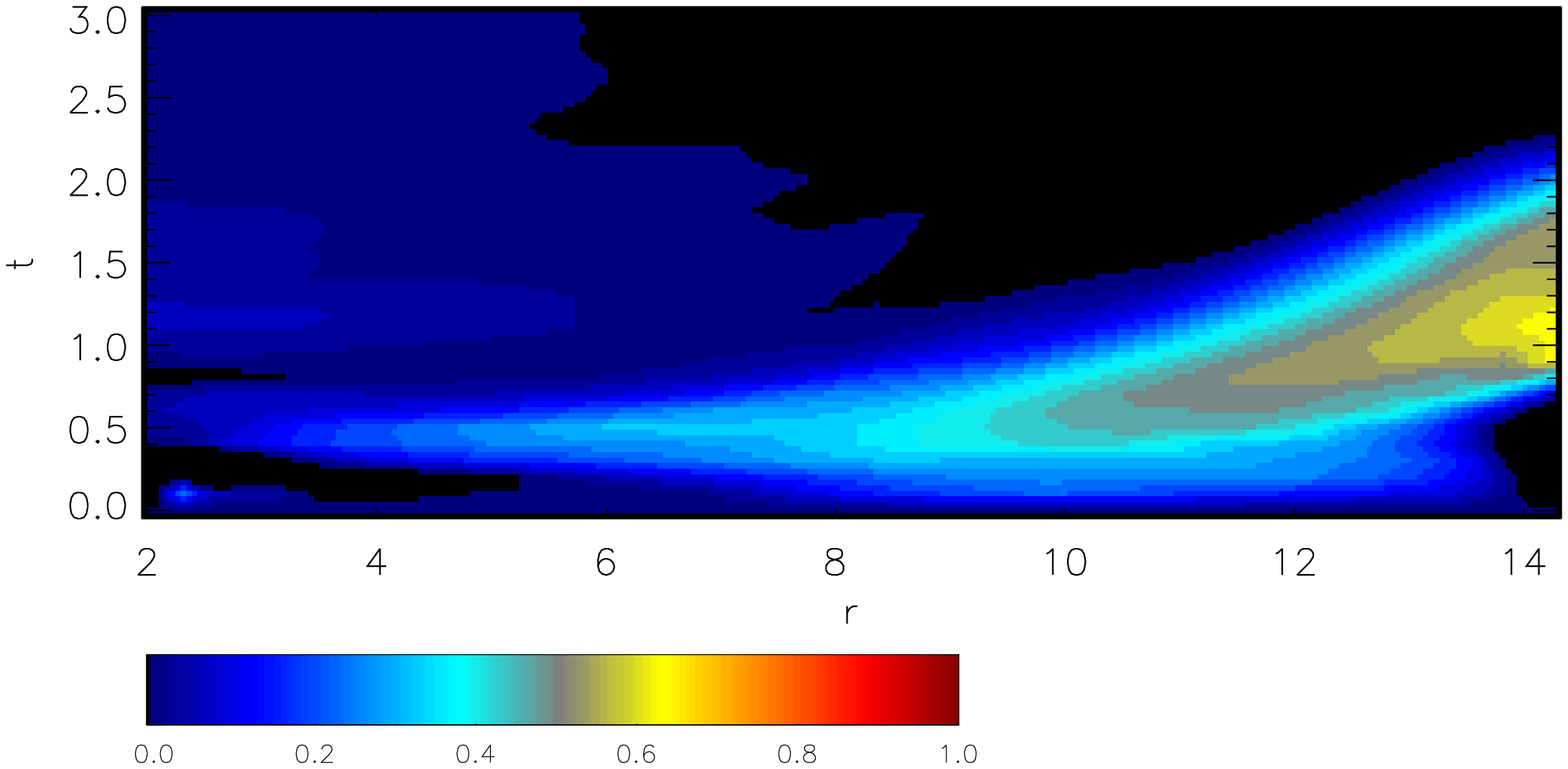}
\includegraphics[width=0.4\textwidth,angle=90]{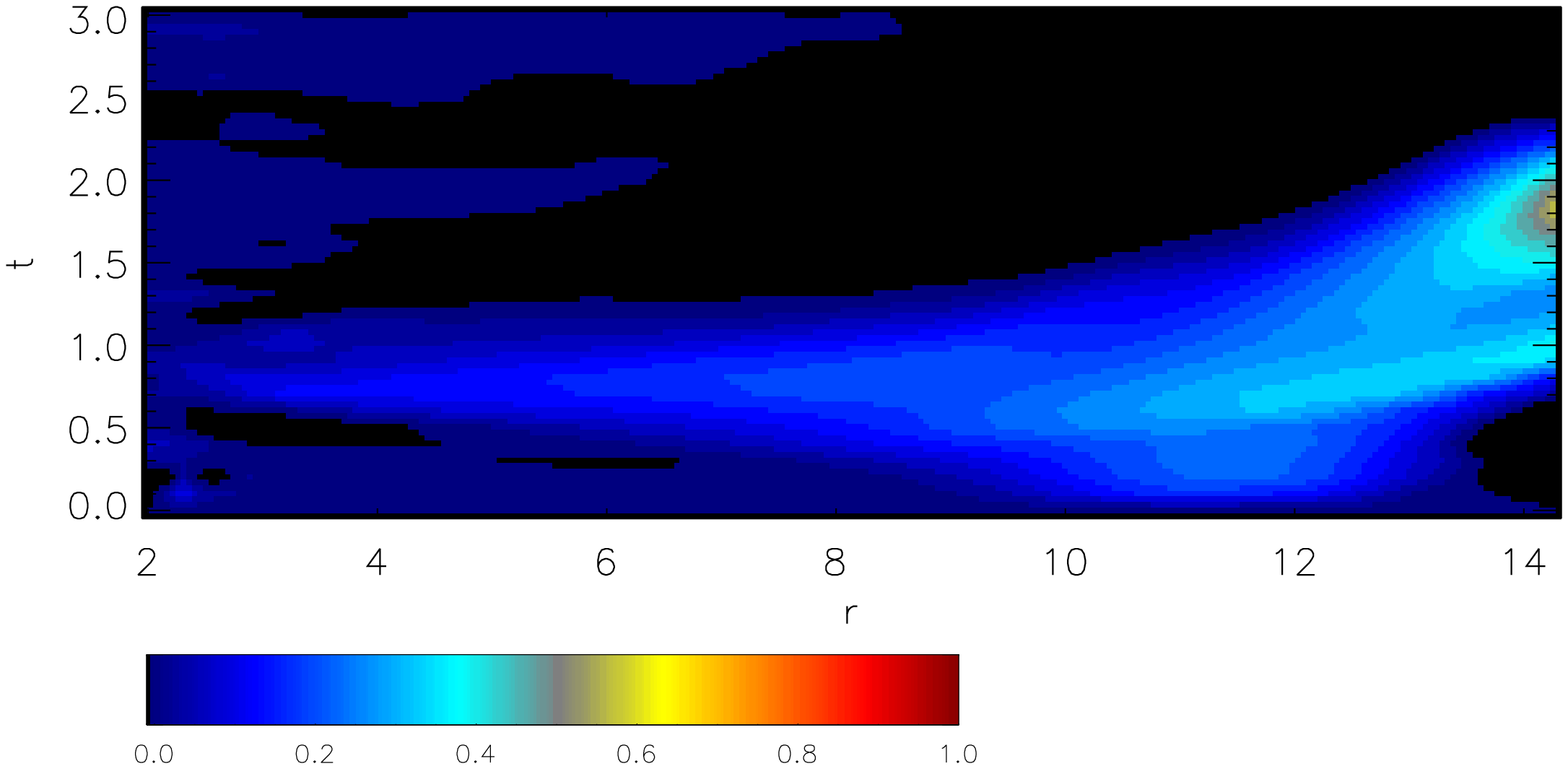}
\caption{The evolution in $r$ and $t$ of the normalized angular momentum flux $S_{rX}/S_0$ in units of the local pressure and radius.  Two simulations sharing the same $c_s$ and $d\theta/d\ln r$, but differing in $L_W$, are shown: \run{2.5W} (top panel) and \run{2.5N} (bottom panel).}
\label{fig:Srx_unequalW}
\end{center}
\end{figure}

We have previously commented on the fact that the exponential decay rate of the warp depends on the warp half-width $L_W$.  This dependence is illustrated in the magnitude of the unaligned angular momentum flux.  In Figure~\ref{fig:Srx_unequalW}, we contrast two simulations with differing warp widths, but identical sound speeds and warping rates $d\theta/d\ln r$, \run{2.5W} and \run{2.5N}.  The peak magnitude of $S_{rX}/S_0$ at $r_c$ increases by a factor $\simeq 2$ as the width of the warp region doubles, a roughly linear scaling.  In other words, the rate at which unaligned angular momentum is transported through the disk depends not only on the local warping rate, but also on the total extent of the warped region.

\begin{figure}
\begin{center}
\includegraphics[width=0.4\textwidth]{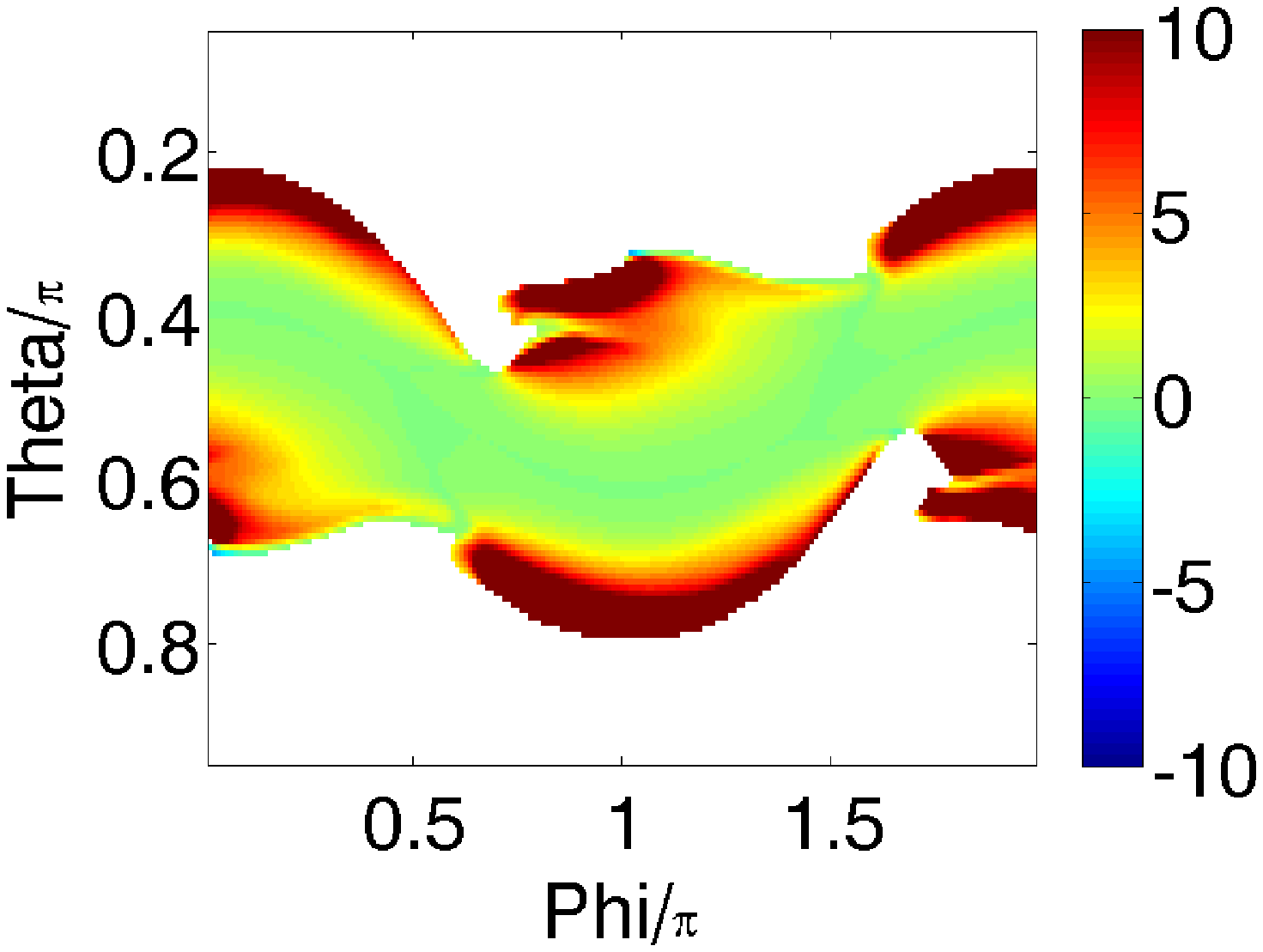}
\caption{The unaligned angular momentum flux $S_{rX}$ in \run{5N} at the middle of the
warp ($r=9$) in the middle of the nonlinear relaxation phase ($t=0.5$~orbits) as a
function of azimuthal angle $\phi$ and polar angle $\theta$ from the axis.}
\label{fig:Srxslice}
\end{center}
\end{figure}

The flux of unaligned angular momentum varies significantly with height within
the disk.  As already shown in Figure~\ref{fig:transonic}, the radial speeds are
greatest on the top and bottom surfaces of the disk.  Although the density is
lower there than in the midplane, Figure~\ref{fig:Srxslice} demonstrates that
the speed of the radial flows is great enough near the surface to compensate
for the lower density, making the flux of unaligned angular momentum also
greatest near the disk surface.  In fact, very little unaligned angular momentum
moves through the central scale height.  The direction of the flows is opposite on
top and bottom, but the sign of $L_X$ is also opposite, making $S_{rX}$ have
the same sign everywhere.  All these patterns are reproduced in the other
simulations as well.  

\begin{figure}
\begin{center}
\includegraphics[width=0.35\textwidth,angle=90]{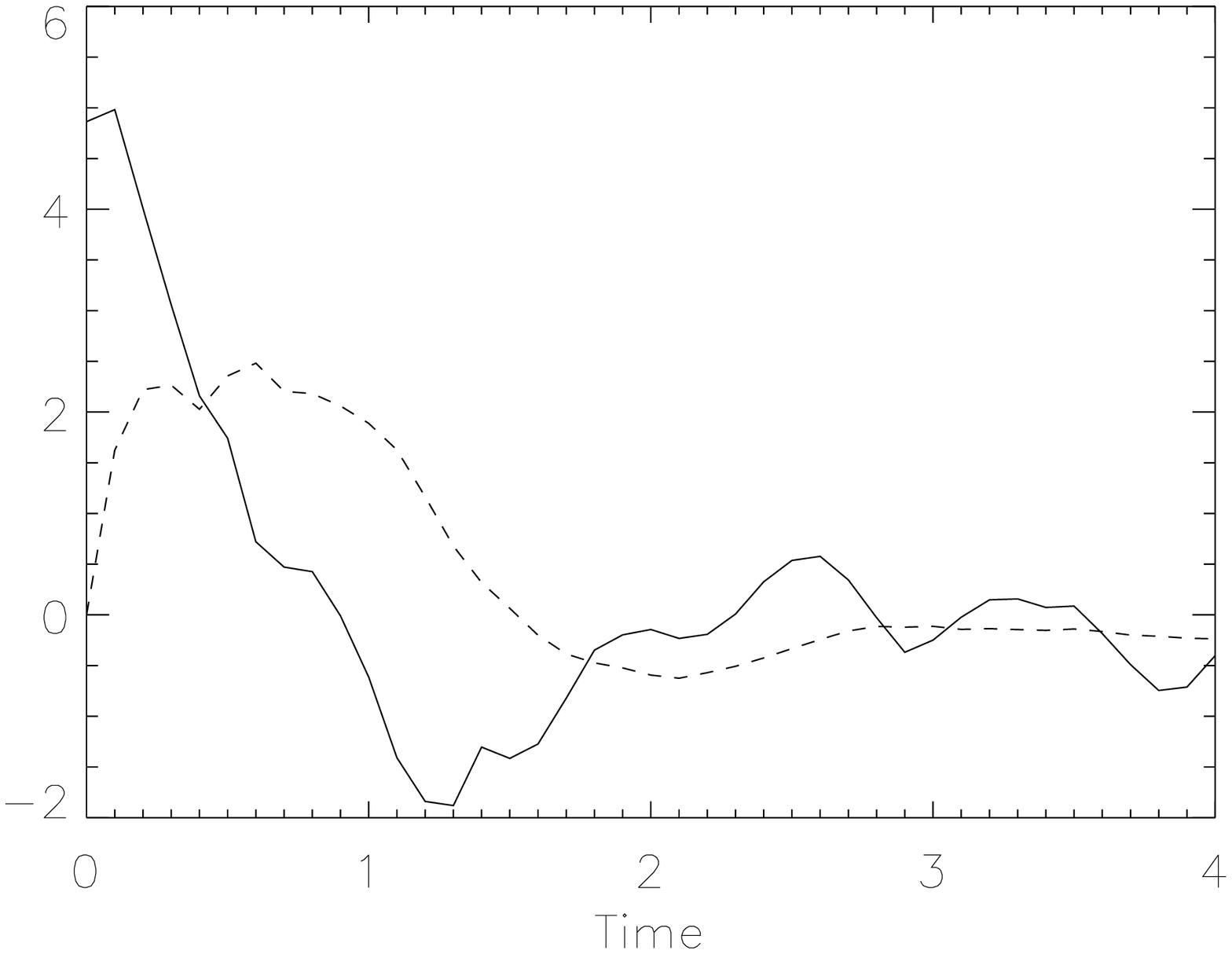}
\includegraphics[width=0.35\textwidth,angle=90]{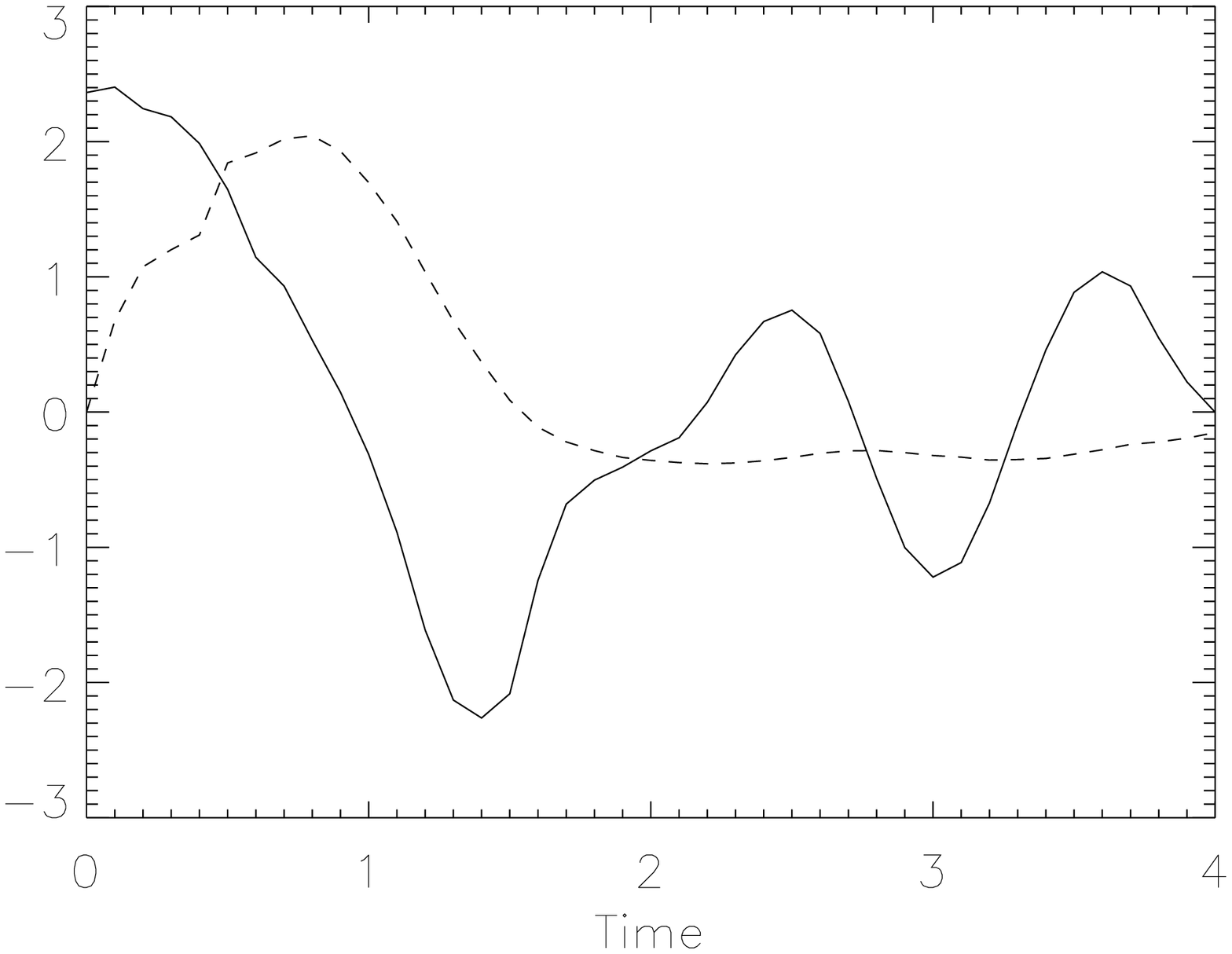}
\includegraphics[width=0.35\textwidth,angle=90]{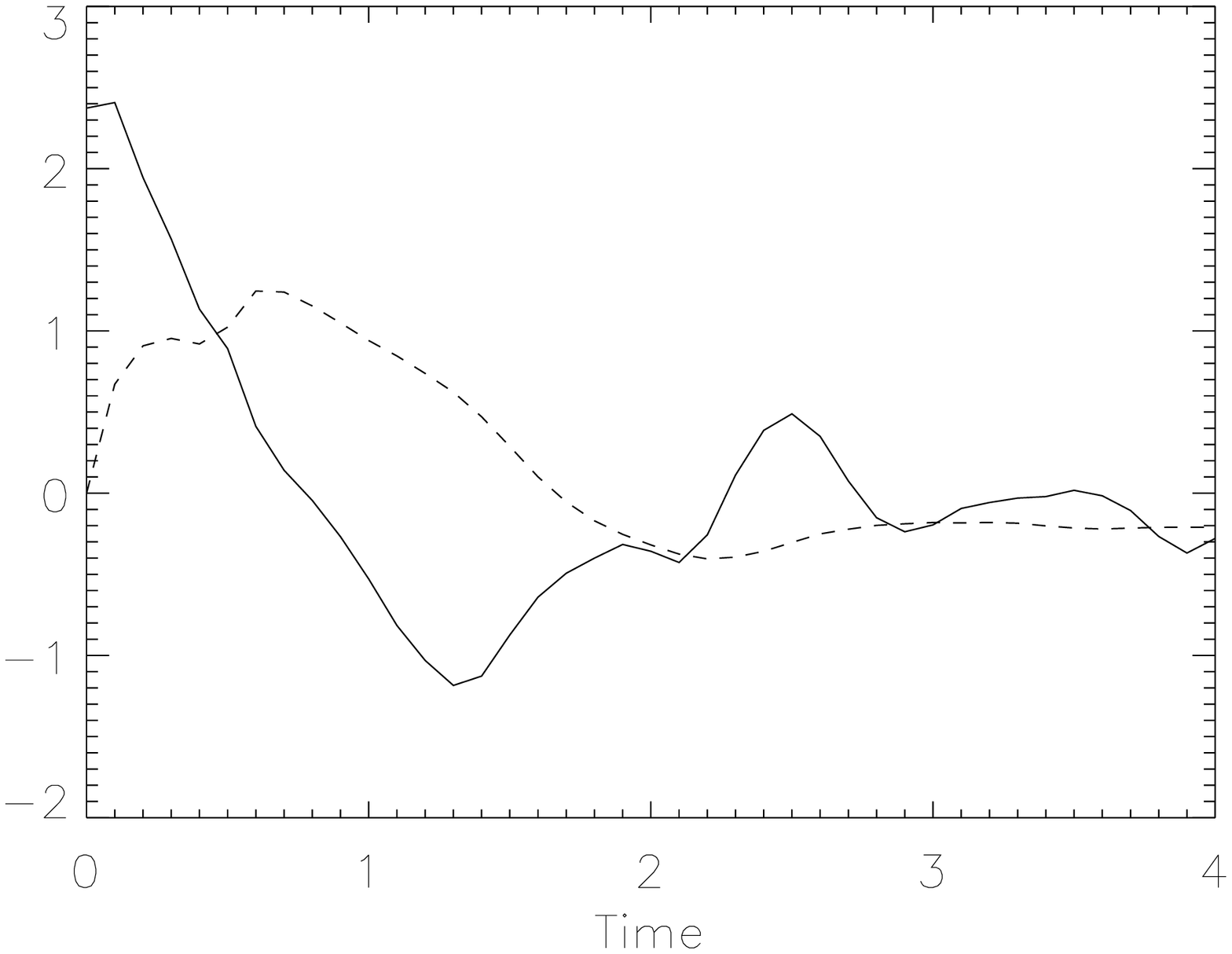}
\includegraphics[width=0.35\textwidth,angle=90]{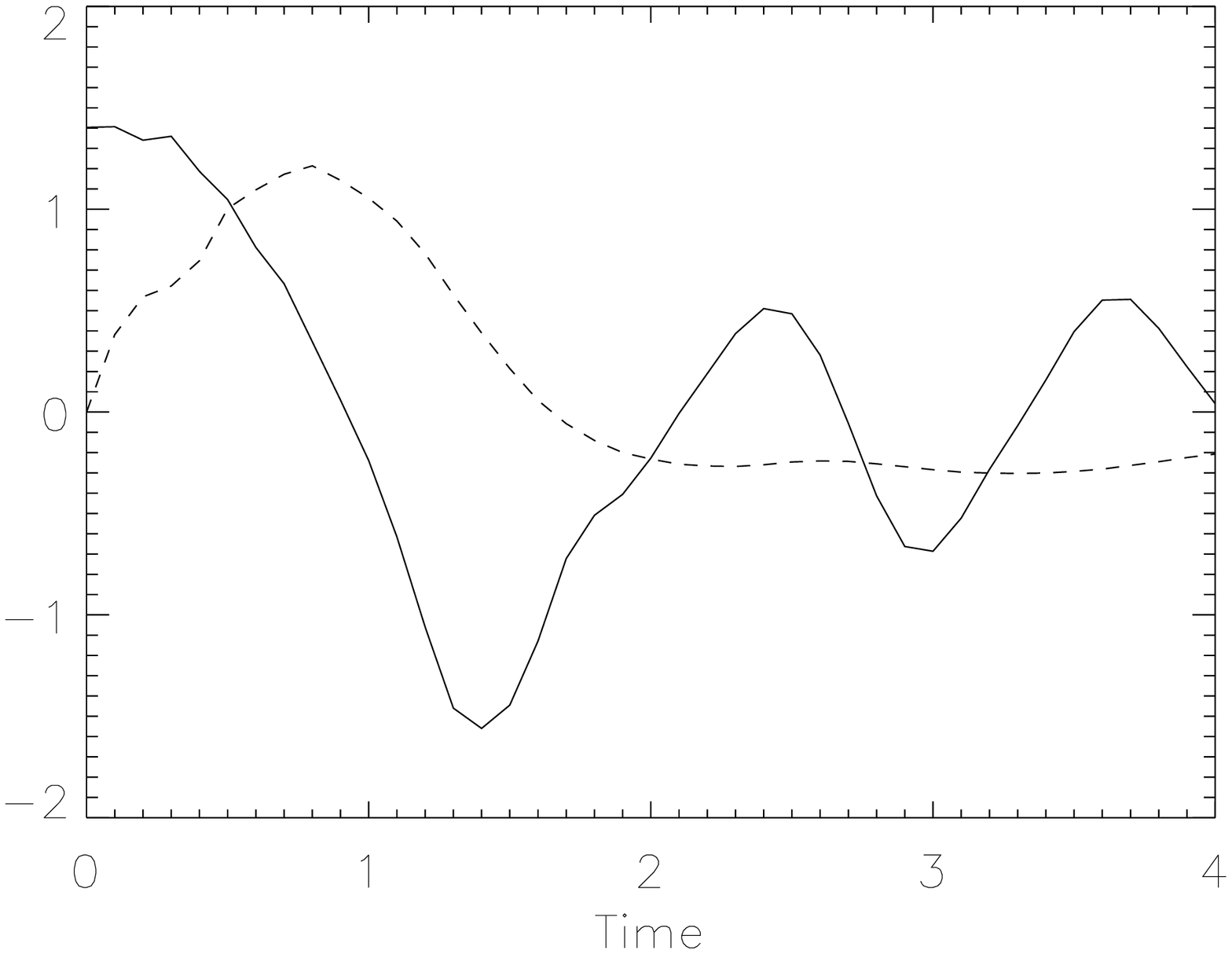}
\includegraphics[width=0.35\textwidth,angle=90]{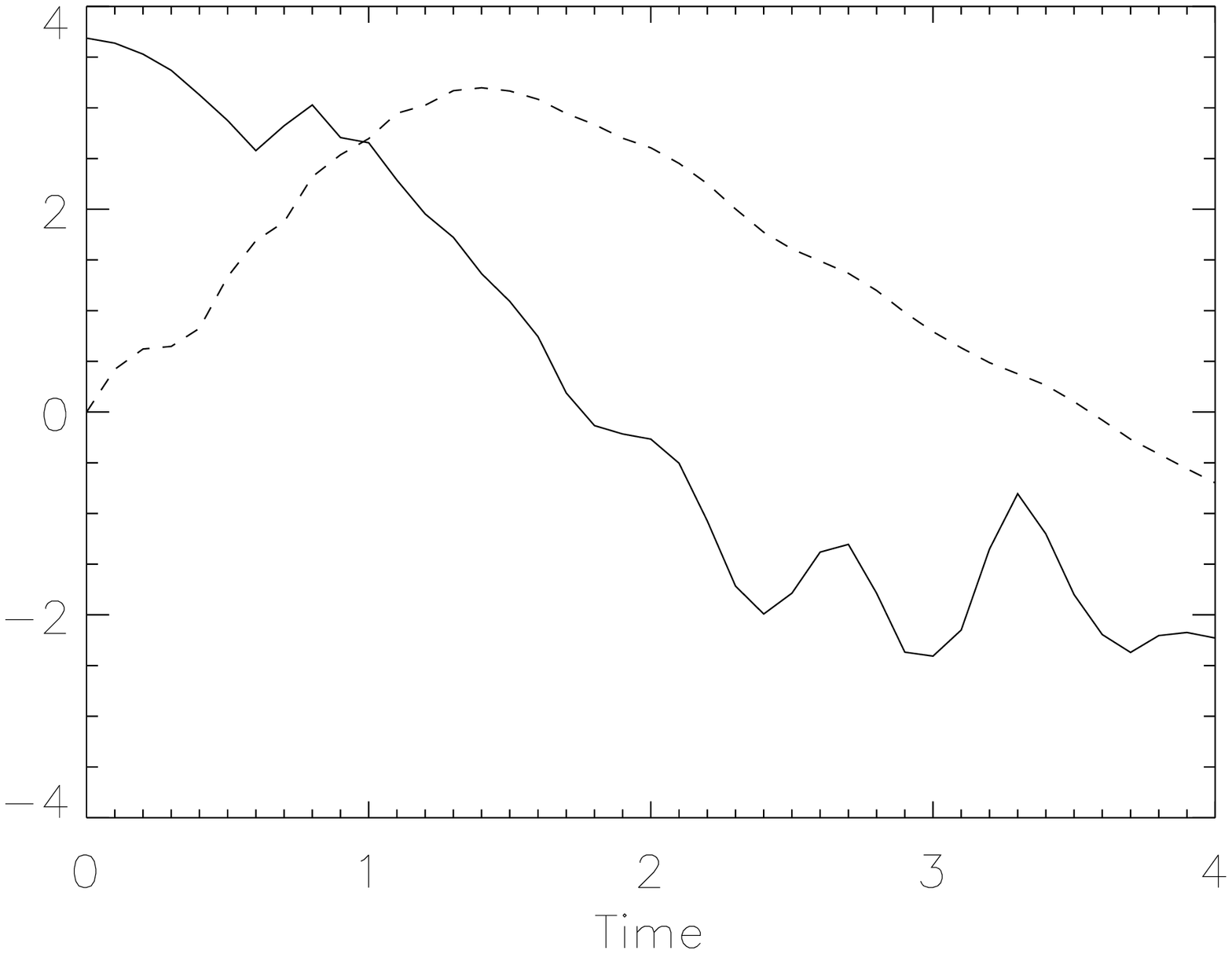}
\caption{Time-dependence of the shell-averaged $\phat$ (solid curve) and
$4S_{rX}/S_0$ at the middle of the warp region in (from left to right and then
top to bottom): \run{HW5N}, \run{2.5W}, \run{2.5N}, \run{1.5W}, and and \runt .
With this arrangement, the two narrow warp width simulations are on the left,
while the wide warp width simulations are on the right and the bottom.
The normalized stress is multiplied by 4 so that its time-dependence can easily be
compared with that of $\phat$. Time is in units of orbits at $r=9$ in the first two
cases, at $r=8.3$ in the third.}
\label{fig:timedep_4orb}
\end{center}
\end{figure}

It is also worthwhile to examine more closely how the unaligned angular momentum
flux varies with time during the relaxation.  That there should be a non-trivial
time-dependence is suggested by the fact that the radial pressure gradients are
determined by the instantaneous warp rate, but pressure gradients determine
instantaneous fluid accelerations, not velocities.  Reynolds stress, which
depends on the velocity, might therefore be expected to lag the pressure
gradient by some characteristic time.  This is indeed the case, as is
shown in Figure~\ref{fig:timedep_4orb}.  In this figure we plot the time-dependence
of both the shell-averaged unaligned angular momentum flux at the
center of the warp and the $\phat$ of the warp shell-averaged at
the same radius.  Time is in orbits at that radius.  During the nonlinear relaxation phase,
the unaligned angular momentum flux $S_{rX}$ noticeably lags the warp rate,
while during the later linear bending wave phase, the lag grows still longer.
For fixed sound speed, the magnitude of this lag depends primarily on the width
of the warped region, not the warp rate.  For example, as shown in
Figure~\ref{fig:timedep_4orb}, the lags in the two runs with $L_W = 2$ (\run{5N}
and \run{2.5N}) are almost identical, both $\simeq 0.6$~orbits, while the
lags in the two runs with $L_W=4$ and the standard sound speed (\run{2.5W}
and \run{1.5W}) are likewise almost identical at $\simeq 0.8$~orbits.
On the other hand, $\runt$, which also has $L_W=4$, but a sound speed
smaller by a factor of 2.5, shows a lag $\simeq 1.4$.
Thus, the characteristic acceleration coherence time appears to be
very roughly given by $\simeq 1.6 (L_W/c_s)^{1/2} \simeq 4 [L_W/(c_s\Omega)]^{1/2}$.
Moreover, this same lag between warp rate and angular momentum flux also explains the
overshoot that creates a persistent linear bending wave after the nonlinear warp rate has
been eliminated.


In those cases in which the time-dependence of $S_{rX}$ approximately follows the time-depence of $\psi$ at a well-defined lag $\tau$, it is of further interest to ask how $S_{rX}(r,t)/S_0$ depends on $\psi(r,t-\tau)$.  We have done so, but find it is rather complex.  At any given radius, there is generally a clear relation between the two in which $S_{rX}/S_0$ increases with increasing $\psi$, but it is typically strongly nonlinear.  In addition, the character of that nonlinearity varies substantially with radius within a single simulation, as well as from case to case.  Thus, these data do not reveal any consistent functional relation between $S_{rX}$ and $\psi$, only that in a rough qualitative sense one rises as the other does.

\subsection{Comparison with Analytic Formalisms}

As we have already remarked, the parameters of our study do not match well with the regimes in which any previous analytic work is valid.  Unfortunately, these previous analytic efforts are the {\it only} ones with which to compare our work, so they are the only options for any attempt to provide some context.  In this section we attempt to situate our work relative to that context.

In past treatments of warp dynamics, the radial mixing of angular momentum by warps has most often been modeled by analogy with diffusion \citep{p92,o99}, even in cases in which $\alpha$ is not large compared to $H/r$ \citep{lp10}.  Two reasons make it worthwhile to contrast our results with those of the diffusion picture, even though our simulations are formally in what is usually called the ``bending wave" regime.  One is that the weakly nonlinear diffusion theory of \cite{o99} was applied to this case, but was found to fail for the specific parameters we study here; comparison with that theory's assumptions may therefore be instructive.  The second is that the physics underlying the diffusion model is in fact not so different from what we have studied.  In both pictures, warps create radial pressure gradients that drive radial flows, and these radial flows mix angular momentum radially.  Where they differ is primarily the mechanism that limits the velocities of these flows and secondarily the nature of the microscopic diffusion that completes the angular momentum mixing.  In the diffusion model, a phenomenological shear viscosity restricts the magnitude of radial motions because their speeds vary with height from the disk plane; in our study, the limit is imposed by a combination of the finite extent of the radial pressure gradient and shocks.  In the diffusion model, that same phenomenological shear viscosity directly mixes momentum; in our case, mixing is achieved by a combination of the localized artificial bulk viscosity mediating shocks and numerical diffusion when velocity gradients are steep even at the grid scale.

As is implied by results we have already discussed, what we observe resembles diffusion in certain ways, but not in all.  Like diffusion, at a single radius at a given time there are fluid elements flowing both in and out, with any net fluxes attributable to the remainder after these flows nearly cancel.  Similarly like diffusion, the end result is to mix conserved quantities, in this case angular momentum.  However, there are also respects in which this process differs from diffusion.  For example, there is comparatively large-scale organization of the regions flowing inward and outward, and the net displacements of fluid elements are not particularly small with respect to the gradient scale.  Nonetheless, for the purposes of mathematical modeling, all that really counts is whether the flux of angular momentum can be described as a diffusion coefficient times the gradient in angular momentum.  More precisely, accounting for the net inward motion of material due solely to the diffusion of the unaligned angular momentum \citep{p92}, a diffusion model predicts that the rate at which unaligned angular momentum passes through the shell at radius $r$ is 
\begin{equation}\label{eqn:pringle_diff}
\pi \nu_2 \Sigma r^2 \Omega \left(\cos\theta + 
     \frac{\sin\theta}{q+2}\frac{\partial\theta}{\partial\ln r}\right)
        \frac{\partial\theta}{\partial\ln r},
\end{equation}
where $\nu_2$ is the diffusion parameter.  With our simulation data, we can test whether the flux is related to the bending rate in this manner.  In particular, the key property that defines a diffusion-like model is whether the flux at a given time and location can be described as the product of a diffusion coefficient (possibly dependent upon the warp) and the warp rate, both evaluated at the same time and location as the flux in question.  For our simulations, the maximum departure angle from the mean angular momentum orientation is $\simeq 0.3$~radians, so the second term inside the parentheses in equation~\ref{eqn:pringle_diff} is small.

Viewing the results reported in the previous subsection from the point of view of this question, several additional contrasts with the diffusion picture appear.  One is that the magnitude of the unaligned angular momentum flux $S_{rX}$ responds not just to the local orientation gradient $d\theta/d\ln r$, but also to the width $L_W$ over which the disk is warped.  Another is that there appears to be a significant delay between the time-dependence of $S_{rX}$ and the time-dependence of $d\theta/d\ln r$, with this delay scaling in rough proportion to $[L_W/(c_s\Omega)]^{1/2}$.  For the parameter range explored here, the magnitude of this delay is $\sim 1$~orbit, comparable to the nonlinear warp relaxation time.  By contrast, in the asymptotic expansion at the heart of \cite{o99}, it is assumed that the shape of the disk changes only on timescales $\sim (H/r)^{-2} \times$ longer than the orbital time. Still another is that although $S_{rX}$ consistently increases with $d\theta/d\ln r$ when the warp rate is evaluated at an appropriately-chosen earlier time, the data do not suggest any consistent functional relation between the two. It is conceivable that the inability of the nonlinear model of \cite{o99} to find a solution for $\alpha_2$ in cases whose parameters are similar to ours (zero anomalous viscosity, epicyclic frequency smaller than the orbital frequency) is related to these disparities.

On the other hand, as we have already remarked, the elimination of any anomalous viscosity from our model may be taken as an indication that the nonlinear warp problem we have treated resembles linear bending waves more than linear warp diffusion.  There are both significant parallels and significant departures between linear bending wave theory \citep{Lubow:2000} and what we observe here.  That the peak stress increases approximately in proportion to the bending rate would be natural for a linear theory.  Similarly, a delay between the bending rate and the torques it induces lies at the heart of bending wave theory.  However, for linear bending waves, the delay timescale defines the period of the oscillations, not the time required for the initiation of damping.  In fact, when $\alpha = 0$ as in our simulations, linear bending wave theory predicts that there is no damping at all.  By contrast, we find that when the bend is nonlinear, it is damped by exactly those Reynolds stresses responsible for wave propagation when the bend is weak.  Further contrasts can be seen in two other facts: that the magnitude of these Reynolds stresses increases in rough proportion to $L_W$, effectively the wavelength of the bending wave; and that the normalized Reynolds stress increases rapidly with decreasing sound speed.  Such nonlinear couplings are, of course, entirely absent in linear theory.

The concentration of $S_{rX}$ near the disk surfaces also underlines the importance of treating properly stresses acting on the vertical shear flow.  Here we have taken an extreme position, assuming zero anomalous viscosity.  Most treatments assume that there is an isotropic anomalous viscosity, one that responds to $r-\theta$ shear as strongly as to $r-\phi$ shear.  Real internal stresses due to MHD turbulence might act in a different way, both in terms of their relationship to shear and in terms of their directional dependence.

\section{Consequences and Conclusions}
\label{sec:con}

We have studied a series of simulations to assess the evolution of strong warps in inviscid disks.  We find that the evolution of these disks is governed by the parameter $\phat$, the radial extent of the warp $L_W$, and the sound speed $c_s$.  A finite thickness flat disk with finite radial extent can achieve hydrostatic equilibrium by pressure gradients with both radial and vertical components: the radial components are balanced by (usually small) departures from Keplerian rotation, while the vertical components are balanced by the vertical component of gravity.  However, a warped disk {\it cannot} be in such an equilibrium; it {\it must} contain unbalanced radial pressure gradients.  As a result, purely acoustic effects stemming from these pressure gradients force radial motions.  When $\phat > 1$, these motions are transonic, mix angular momentum radially, and rapidly relax the warp.

In our simulations, conducted with neither any explicit viscosity nor the sort of MHD stresses that convey angular momentum in a conventional flat disk, the speeds of the radial motions depend on the overall strength of the warp, a combination of $\meanpsi$ and the radial extent over which the bending takes place.  In the parameter regime we studied, the nonlinear exponential damping rate $s \simeq 0.14 (\psi/0.6)^{b}(L_W/r)\Omega$ with $b \simeq 1$--1.5, and is almost independent of $c_s$.  When we changed the sound speed at fixed $\psi$, the Mach number of the radial motions changed in the opposite direction, leaving the radial speed largely unaltered.  It is possible that the Mach number saturates for still larger values of $\phat$, so that $s$ declines with decreasing $c_s$, but in the parameter range covered by our simulations no such trend was apparent.
Even though the fluid motions are only transonic, in disks like the ones we have studied, where $H/r$ is not $\ll 1$, it is possible for streams to cross the entire span of the warp in roughly an orbital period because $(r/c_s)(\Omega/2\pi) \simeq (1/2\pi)(H/r)$.  As a result, in our simulations relaxation from a state of strongly nonlinear warping to linear behavior required only a few mid-warp orbits.

Relaxation of warps in a purely hydrodynamic disk requires fluid motions to mix
angular momentum from regions initially having different orientations.  The flux
of unaligned momentum in the radial direction, a quantity we dub $S_{rX}$, therefore
becomes the key quantity governing the relaxation.  Scaling this quantity in units
of the product of the local pressure and radius, we find that $S_{rX}$ increases
with warp rate, but also with warp {\it width} $L_W$; in other words, it is controlled
in part by global conditions, not local.  Moreover, as might be expected when the
fluid motions carrying angular momentum arise from the acceleration in pressure
gradients directly associated with the warp, $S_{rX}$ varies in a way that is
related to the warp rate, but delayed by a lag that scales $\propto (L_W/c_s)^{1/2}$.
For the parameters of our simulations, these lags were comparable to the decay time;
in cooler, thinner disks, this scaling suggests that they become even longer.

These results suggest that decay of nonlinear warps conforms to neither a conventional
diffusion nor a bending wave formalism.  The process depends on
a net divergence in unaligned angular momentum flux, as in a diffusion picture,
but this flux is determined both by the history of the warp (because of the significant
delay between changes in the warp and the creation of angular momentum fluxes) and global
disk properties.  On the other hand, bending
wave dynamics do not entirely suffice because the same Reynolds stresses that lead to
bending wave propagation in the linear regime create rapid warp relaxation, i.e., wave
damping, when they become nonlinear.  Moreover, no anomalous viscosity is required to
control the speed of the radial motions; purely hydrodynamic effects, such as the
inability of thermal pressure gradients to accelerate fluid to more than a few times
the sound speed, and the creation of weak shocks, suffice.

The simulations presented here are by no means the first computational study performed of warped disks.  There has been previous work \citep{np99,lp07,lp10} using SPH simulations to study this phenomenon.  However, our simulations are both the first to use a global grid-based treatment and the first to focus on hydrodynamics without any anomalous viscosity.  \citet{np99} studied the propagation of bending waves excited by pulses of varying amplitudes placed in the middle of a disk whose imposed anomalous viscosity was small compared to the disk thickness ($\alpha < H/r$).  Like them, we found strong damping of nonlinear amplitude disturbances by radial motions, but they did not study how the relaxation rate or its associated angular momentum fluxes depended on parameters.  More recent SPH work \citep{lp07,lp10} has focused on the ``diffusive'' regime, in which $\alpha > H/r$; even the simulations they labeled as ``inviscid" assumed an $\alpha \simeq 2 H/r$.  They are therefore not directly comparable to our work.  In addition, like the earlier simulational efforts, they did relatively little in the way of parameter exploration or investigation of the detailed time-dependence of warp relaxation, so we cannot compare our results in those regards to theirs.  Nonetheless, it is interesting to note that \cite{lp10} speculated that when $\alpha$ is small and $d\theta/d\ln r$ comparatively large (i.e., in our circumstances), a diffusion model might be inappropriate.
  
%

The absence of ordinary internal stresses (i.e., those produced by MHD turbulence or often modeled by an anomalous viscosity) might be considered a limitation of this work.  We chose to eliminate them in order to clarify which processes account for which effects.  In future work on this subject we will include explicit MHD stresses in order to see what new effects they introduce.  In particular, it will be especially important to see the degree to which MHD turbulence restricts the growth of the transonic radial motions crucial to warp relaxation, and what relation that degree of restriction has to the stresses responsible for angular momentum transport in flat disks.  As we have shown, the concentration of unaligned angular momentum flux to the surfaces of the disk is so strong that careful treatment of any stresses induced by this shear could be quite important.  It is possible, however, that the relevant MHD stress (the $r-\theta$ component) will remain a small effect.  In most simulations of MHD turbulence stirred by the magnetorotational instability from \cite{shgb96} onward, $B_\theta$ is the smallest of the three field components, only $\sim 0.1 B_\phi$.  Consequently, even at distances one or two scale heights off the midplane, where the shear is strongest, the MHD stress restraining the radial motions may be weak.

\section*{Acknowledgements}
We would like to thank Steve Lubow for useful discussions and advice.  This work was partially supported under National Science Foundation grants AST-1028111 and AST-0908326 (JHK and KAS), AST-0908869 (JFH), and NASA grant NNX09AD14G (JFH).  Computational resources for this project were provided by the Homewood High Performance Compute Cluster supported by the NSF through grant NSF-OCI-108849.  

\appendix
\section{Numerical Tests}
\label{sec:num}

The rotational invariance of the Euler equations implies that the simulation results should be invariant under a rigid rotation.  Although this is true physically, this invariance will not, in general, be respected numerically.  In special cases, when the velocity is parallel to a symmetry direction of the flow and this direction is along a coordinate axis, fluid transport creates very little numerical dissipation.  A statistically axisymmetric orbiting disk treated in either cylindrical polar or spherical coordinates whose equatorial plane coincides with the disk plane is an example of this favorable special case.  When that same disk is inclined to the equatorial plane, however, the numerical dissipation can be considerably enhanced.  Unfortunately, when a disk's inclination changes from place to place, it is impossible to avoid having the dominant velocity oblique to the coordinate system somewhere, yet understanding the physics associated with a warped disk relies on being able to separate physical effects from numerical ones.  We therefore took special pains to choose a combination of coordinate system, inclination, and grid resolution that would keep numerical dissipation small enough to be negligible.

Our procedure was to consider a series of disks with a constant, but non-zero, inclination angle in both cylindrical and spherical coordinates at both low and moderate resolution, roughly 4~ZPH and 8~ZPH respectively (ZPH is zones per scale height).  Both grids employed isotropic cells.  These flat, but tilted, disks had the same radial extent and thickness as the production run initial conditions.  Each simulation was run for $10$ orbits at $r=r_c$.  To quantify numerical losses, we computed two measures: the fractional change in the magnitude of the fluid's total angular momentum $\Delta |\vec L|/|
\vec L|$ and the fractional change in the density-weighted tilt $\theta_{T}$.  The latter quantity is defined as
\begin{equation}
\label{eqn:thetatilt}
\theta_{T}(t) = \tan^{-1} \left( \frac{-<L_{x}>}{<L_{z}>} \right).
\end{equation}  
Here the components of the angular momentum $\vec{L}$ are taken in the grid coordinate system.  The results from this study are shown in Table~\ref{tab:oblique}.  Note that the net change in the magnitude of the angular momentum is {\it always} negative, and that any change in mean inclination is always in the sense of bringing the disk closer to the coordinate system's equatorial plane.

\begin{table}
\begin{center}
  \begin{tabular}{@{} |c|c|c| @{}}
    \hline
    Run ID & -$\Delta L/L$ & -$\Delta\theta_T/\theta_T$ \\ 
    \hline
C15L & 1.45\% & 5.46\%  \\ 
C15M & 0.63\% & 1.83\%  \\ 
C30L & 3.24\% & 6.76\%  \\ 
C45L & 11.08\% & 7.44\%  \\ 
C45M & 3.51\% & 1.99\%  \\ 
C60L & 23.63\% & 9.20\%  \\ 
S15L & 0.54\% & 2.58\%  \\ 
S15M & 0.41\% & 1.07\%  \\ 
S30L & 1.49\% & 3.20\%  \\ 
S30M & 0.71\% & 1.12\% \\
S45L & 8.56\% & 7.28\%  \\ 
S45M & 3.29\% & 2.46\%  \\ 
S60L & 24.61\% & 12.74\%  \\ 
     \hline
  \end{tabular}
\end{center}
\caption{Results from a series of runs to test the numerical importance of disk inclination.  Simulations are denoted by the geometry used: (C)ylindrical or (S)pherical; the angle (in degrees) of the oblique inclination; and whether the resolution was (L)ow or (M)oderate, using $4$ or $8$ zones for each vertical scaleheight.}
\label{tab:oblique}
\end{table}

Comparing cylindrical and spherical tests at the same $\theta_T$, we find that spherical coordinates are clearly better according to both measures when $\theta_T < \indeg{45}$.  At $\theta_T = \indeg{45}$, the verdict is mixed: spherical coordinates are better by the $\Delta L/L$ criterion, cylindrical by the $\Delta\theta_T/\theta_T$ criterion.  At larger angles, both coordinate systems do poorly, and to a similar degree.  The results for low inclination angles, however, give us confidence that the study of warped disks in which the inclination angle doesn't exceed the threshold of $\indeg{45}$ can be conducted with confidence, provided the resolution is at least 8~ZPH.  For our production runs, we used exclusively spherical coordinates due to their superior performance for the inclination angles of interest.  In particular we note that our warped simulations are most closely comparable to the oblique disk simulation S30M, for which the fractional losses in both the angular momentum and inclination angle are less than or approximately $1\%$.

\begin{figure}
\begin{center}
\includegraphics[width=0.8\textwidth]{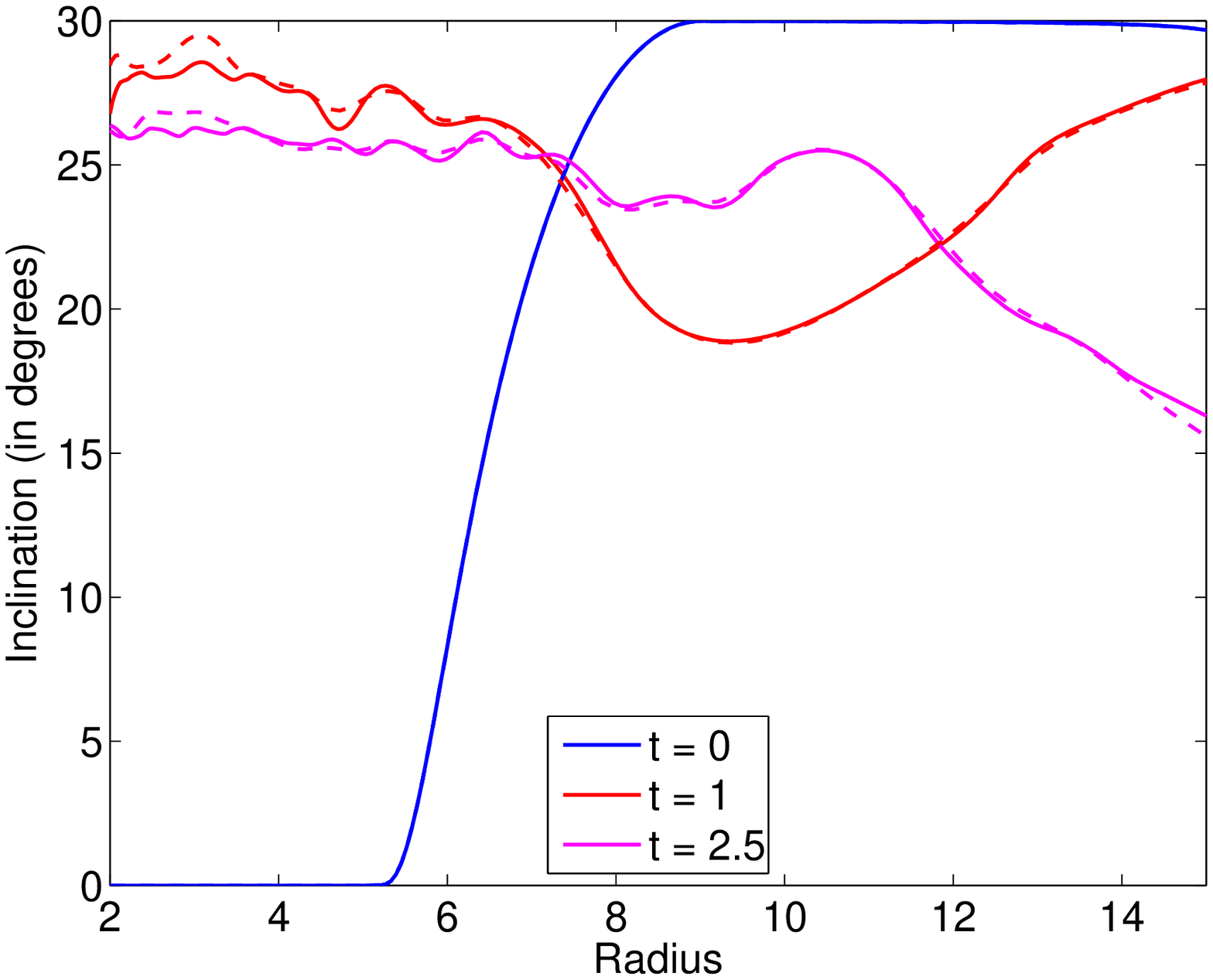}
\caption{Resolution test of a warped disk.  Shown are the radial inclination profiles at various times for two simulations differing only in their resolution (8~ZPH: dashed, 16~ZPH: solid). }
\label{fig:conv}
\end{center}
\end{figure}

There exists much work in the literature discussing criteria for the proper resolution of an accretion disk simulation.  Although satisfying these constraints will be particularly important when internal stresses studied because they are caused by MHD turbulence, for pure hydrodynamics the resolution constraints are much weaker.  Nonetheless, we tested the resolution-dependence of our results.  We conducted two simulations identical save for their resolution, which corresponded to roughly eight and sixteen zones per vertical scale height.  Snapshots of the evolution with time of their inclination profiles, $\theta_{T}(r,t)$ are shown in Figure~\ref{fig:conv}, where
\begin{equation}
\label{eqn:thetatiltst}
\theta_{T}(r,t) = \tan^{-1} \left( \frac{-<L_{x}>_{\theta \phi}}{<L_{z}>_{\theta \phi}} \right),
\end{equation}
and the notation $X(r) = <X>_{\theta \phi}$ is used to denote a shell average of a quantity $X$.  Overall, we find excellent agreement, and in particular find almost indistinguishable behavior within the body of the disk ($2<r<15$).  The comparison shown here is limited to $2.5$ orbits at $r=r_c$, but, as we will show, it is during this period that the most violent realignment of this disk occurs.  This is therefore the most important time to examine.

\bibliographystyle{apj}
\bibliography{Bib}
\end{document}